\documentclass[preprint,superscriptaddress,nofootinbib,tightenlines]{revtex4}

\usepackage{epsfig}

\def\OMIT#1{{}}

\def\si{^1 \hskip -0.03in S _0}
\def\siii{^3 \hskip -0.025in S _1}

\newcommand{\lsim}{\raisebox{-0.7ex}{$\stackrel{\textstyle <}{\sim}$ }}
\def\pislash{ {\pi\hskip-0.54em /} }

\def\nopi{ {\rm EFT}(\pislash) }

\begin{document}

\preprint{\vbox{
\hbox{UNH-03-01}
\hbox{LBNL-53972}
\hbox{NT@UW-03-031}
}}

\vphantom{}

\title{Exploring Hyperons and Hypernuclei with Lattice QCD}

\author{ S.R.~Beane}
\affiliation{Department of Physics, University of New Hampshire,
Durham, NH 03824-3568.}
\affiliation{Jefferson Laboratory, 12000 Jefferson Avenue, 
Newport News, VA 23606.}
\author{P.F.~Bedaque}
\affiliation{Lawrence-Berkeley Laboratory, Berkeley,
CA 94720.}
\author{A.~Parre{\~n}o}
\affiliation{Dept. ECM, Facultat de F\'{\i}sica, Universitat de Barcelona,
E-08028, Barcelona, Spain.}
\author{M.J.~Savage}
\affiliation{Department of Physics, University of Washington, 
Seattle, WA 98195-1560.\\
\qquad}

\vphantom{}
\vskip 0.5cm
\begin{abstract} 
\vskip 0.5cm
\noindent In this work we outline a program for lattice QCD that would
provide a first step toward understanding the strong and weak
interactions of strange baryons.  The study of hypernuclear physics
has provided a significant amount of information regarding the
structure and weak decays of light nuclei containing one or two
$\Lambda$'s, and $\Sigma$'s.  From a theoretical standpoint, little is
known about the hyperon-nucleon interaction, which is required input
for systematic calculations of hypernuclear structure.  Furthermore,
the long-standing discrepancies in the $P-$wave amplitudes for
nonleptonic hyperon decays remain to be understood, and their
resolution is central to a better understanding of the weak decays of
hypernuclei.  We present a framework that utilizes L{\"u}scher's
finite-volume techniques in lattice QCD to extract the scattering
length and effective range for $\Lambda N$ scattering in both QCD and
partially-quenched QCD.  The effective theory describing the
nonleptonic decays of hyperons using isospin symmetry alone,
appropriate for lattice calculations, is constructed.
\end{abstract}

\maketitle

\vfill\eject

\section{Introduction}

\noindent Nuclear physics is a fascinating field that remains poorly
understood at a fundamental level.  While the underlying Lagrange
density describing all strong interactions is well established to be
Quantum Chromo Dynamics (QCD), its solution in the nonperturbative
regime leads to many unexpected and intriguing phenomena when the
fundamental constants of nature take their physical values.
Furthermore, one knows effectively nothing about how most strongly
interacting systems behave as the fundamental constants are allowed to
move away from their physical values. A puzzling feature of nuclear
physics is the unnaturally large size of the $S-$wave nucleon-nucleon
scattering lengths; i.e.  one would like to understand why low-energy
QCD has chosen to be very near an infrared unstable fixed point of the
renormalization group, as this requires a non-trivial conspiracy
between the quark masses and the scale of QCD, $\Lambda_{\rm QCD}$.
While nuclei composed of neutrons and protons are well studied
throughout most of the periodic table, and further efforts are
proposed to study very short-lived nuclei that impact stellar
environments, relatively little is understood about nuclei that
contain one or more valence strange quarks: the hypernuclei.
Understanding of the structure and decays of hypernuclei from first
principles remains at a primitive level, and it is possible that the
study of hypernuclei may help in unraveling fundamental issues
regarding ordinary nuclei in terms of QCD.

Since the $\Lambda$ hyperon is the lightest among the hyperons
($m_\Lambda \sim 1115~{\rm MeV}$), most theoretical and experimental
efforts have concentrated on bound systems of non-strange baryons and
one $\Lambda$. These single-$\Lambda$ hypernuclei can be
experimentally produced by strangeness-exchange reactions
$N(K,\pi)\Lambda$ at CERN, BNL, KEK and Da$\Phi$ne, by strangeness
associated production reactions $n(\pi, K)\Lambda$ at BNL and KEK, or
by the electroproduction mechanism $p(e,e' K^+ \;) \Lambda$ at JLab
(also planned at GSI for 2006)~\cite{exp}.  These reactions usually
populate highly-excited hypernuclear states, which decay
electromagnetically to leave the hyperon in the lowest energy state,
$1s_{1/2}$. Once in the ground state, the system decays weakly,
without conserving strangeness, parity or isospin. For very-light
nuclei, such as $^3_\Lambda$H, the decay proceeds mainly through the
mesonic decay mode, $\Lambda \to N \pi$, which is analogous to the
decay of the $\Lambda$ hyperon in free space.  These pionic decay
channels are highly suppressed when the $\Lambda$ is embedded in a
nuclear medium, as the momentum of the outgoing nucleon ($\sim
100~{\rm MeV}$) is small compared to the typical Fermi momentum ($\sim
270~{\rm MeV}$ for infinite nuclear matter). Non-mesonic
(multi)nucleon induced decay channels, mainly $\Lambda N \to NN$, are
then responsible for the decay of the $\Lambda$ hyperon in the medium,
and become dominant for p-shell and heavier systems (note that even
for $^5_\Lambda$He, the non-mesonic decay channel amounts to
$40\%-50\%$ of the total decay rate~\cite{heliexp}).  The lack of
stable hyperon beams (the $\Lambda$ lifetime is $\tau_\Lambda \sim 2.6
\times 10^{-10}~{\rm s}$) makes it difficult to perform a study of
the strong and weak hyperon-nucleon (YN) interaction in free
space. Therefore, hypernuclear structure and decay experiments
presently provide the only quantitative information on the YN
interaction.

Experimental information on hypernuclear decay results from the measure
of total lifetimes, partial decay rates (typically from the detection
of a proton in the final state, and more recently from the detection
of $nn$ and $np$ pairs in coincidence) and angular asymmetry in the
emission of protons from the decay of polarized hypernuclei. However,
the existing data have large error bars, in part due to poor
statistics. On the theoretical
side, one has to fold a particular model for the weak $\Lambda N \to
NN$ reaction into a many-body system.  Realistic ground-state wave
functions in the $S=-1$ sector have been derived for $A \le 4$
hypernuclei using the Faddeev-Yakubovsky
formalism~\cite{hypertriton,nogga}, and up to $p-$shell hypernuclei
using a three- and four-body cluster structure solved with the
Gaussian-basis coupled-rearrangement-channel method~\cite{hiyama}.
The complete description of the final few-body scattering states in
hypernuclear decay has only been performed for $A=3$ using the Faddeev
formalism~\cite{hypertriton}.  For heavier systems, finite nuclei are
typically studied within a shell-model framework, while the final
state is treated as a distorted wave function describing the two
weakly-emitted nucleons moving away from a residual $(A-2)$ core. 
In any of those calculations, a model for the YN interaction
(typically a one-meson-exchange model) is assumed in order to describe
the strong two-body dynamics. The development of
Effective Field Theories (EFT) in the strong $S=0$ sector, motivated
recent model-independent work in the $S=-1$ sector. 
An EFT~\cite{rob} was constructed at next-to-leading order (NLO)
for momenta below pion-production threshold, 
and used to study scattering YN observables,
as well as hyperon mass shifts in the medium, via a low-density
expansion. The existence of the hypertriton was used to constrain
the $\Lambda N$ singlet- and triplet-scattering lengths.
However,  a
description of bound states using this systematic two-body formalism 
does not yet exist. 

Regarding the $|\Delta S|=1 \;\; \Lambda N$ interaction, all the
models developed to date agree in describing the long-range part of the weak
transition through the exchange of Goldstone bosons, $\pi, K$ and
$\eta$\footnote{The $\eta$-exchange mechanism is omitted in some works
due to the smallness (a factor of 10) of the $\eta NN$ coupling
constants.}. For the short-range part of the interaction, some studies
adopted the exchange of heavier mesons~\cite{DFHT96,OMEW}, the vector
$\rho, \omega$ and $K^*$ mesons, while others used 
a quark-model type description~\cite{ISO00}.  
Although some progress has been achieved
during the last ten years, none of the existing models give a
satisfactory explanation of the decay process. Construction
of a model-independent EFT description of these decays
was initiated in Refs.~\cite{Parreno:2003ny,Parreno:2003mf}.  
The
study of the $\Delta S=1 \;\; \Lambda N$ interaction has the
additional charm of allowing access to both the
parity-conserving (PC) and the parity-violating (PV) part of the weak
interaction, in contrast to its $\Delta S=0 \;\; NN$ partner, 
where
the PC weak signal is masked by the strong PC amplitude which is
10 orders of magnitude larger.

A major tool for strong-interaction physics that will play an
ever-increasing role in nuclear physics is lattice QCD.
First-principles simulations on the lattice are the only known way to
perform rigorous calculations of observables and processes where the
QCD strong interaction contributes.  Present-day calculations in
lattice QCD are restricted to fairly large quark masses, relatively
small volumes and relatively large lattice spacings.  However, as time
progresses the lattice-quark masses are decreasing, lattice sizes are
increasing, and lattice spacings are decreasing. Furthermore, it has
been appreciated that unphysical simulations performed with
partially-quenched QCD (PQQCD) allow for the determination of
unquenched QCD observables, and can be performed sooner than fully
unquenched simulations. While significant effort has been placed in
calculating observables in the meson sector with impressive success,
systems involving more than one baryon have largely been ignored with
the exception of one pioneering effort to compute the nucleon-nucleon
scattering length in quenched QCD~\cite{Fukugita}.

In this work we establish the framework with which lattice (PQ)QCD can
determine basic strong-interaction parameters important for
hypernuclei: the scattering parameters for $\Lambda
N\rightarrow\Lambda N$.  Given the presently large experimental
uncertainties in this scattering amplitude, a lattice calculation will
provide an important contribution to refining the potentials that are
input into hypernuclear-structure calculations.  Another component of
this work is to explore the fundamental ingredients of calculations of
weak decays of hyperons that must be understood before we are able to
claim an understanding of the weak decays of hypernuclei
themselves. Such an understanding has eluded theorists for decades;
while one appears able to describe the $S-$wave amplitudes in
the hyperon nonleptonic weak decays, the $P-$wave amplitudes are very-poorly 
predicted. 
Lattice QCD is perhaps the only hope for gaining
an understanding of these fundamentally-important weak processes.

\section{Strong $\Lambda N$ Scattering}

\noindent In predicting the structure of hypernuclei, the
hyperon-nucleon potential or the hyperon-nucleon scattering amplitude
is required input.  A compilation of the world's experimental data on
elastic hyperon-nucleon scattering can be found at the Nijmegen
``NN-Online'' website~\cite{NNonline}. The cross-sections in
particular energy regions for these processes have large uncertainties
associated with them as the total number of events is quite small~\footnote{We are grateful to Rob Timmermans for
discussions about the state of hyperon-nucleon scattering data.}.
To get a sense of the uncertainty in the effective range parameters,
consider that Ref.~\cite{Sechi-Zorn:hk} quotes the following one-standard-deviation bounds
for the $S-$wave scattering lengths and effective ranges:
\begin{eqnarray}
0.0&>&a^{(\si)}>-15~{\rm fm}\quad\qquad 0.0>r^{(\si)}>15~{\rm fm}\nonumber \\ 
-0.6&>&a^{(\siii)}>-3.2~{\rm fm}\;\;\qquad 2.5>r^{(\siii)}>15~{\rm fm} \ \ .
\label{eq:lambdaNexpert}
\end{eqnarray}
By comparison, a contemporary study~\cite{Alexander:cx} finds (best fit) 
$a^{(\si)}=-1.8~{\rm fm}$, $r^{(\si)}=2.8~{\rm fm}$, and
$a^{(\siii)}=-1.8~{\rm fm}$, $r^{(\siii)}=3.3~{\rm fm}$.
Unfortunately, the meaning of these effective-range analyses is
unclear.  They arise from a four-parameter fit to data with very-poor
statistics at energies where one would not necessarily expect shape
parameters and other higher-order terms in the effective-range
expansion to be small. Moreover, the effective-range parameters fit to
the data are highly correlated.  While the bounds of
eq.~(\ref{eq:lambdaNexpert}) are therefore suspect, they do encompass many
model-dependent analyses of hyperfragments, production reactions and
potential models~\cite{hyperexamples}.  What then is known? Given the
current state of experiment, it is probably safe to say: (i) there is
no hyperdeuteron and therefore $a^{(\si)}<0$ and $a^{(\siii)}<0$.
(ii) consistency of potential models with the spin of the hypertriton
implies $|a^{(\si)}|>|a^{(\siii)}|$.

With the lack of experimental effort in the direction of
Lambda-nucleon scattering, lattice QCD may provide a rigorous,
first-principles calculation of these cross-sections over some energy
interval with smaller uncertainties than those determined
experimentally in the near (and possibly far) future. In this section
we put in place the theoretical framework required to connect possible
future lattice calculations with data in the low-energy regime, below
the pion-production threshold.

The past decade has seen remarkable progress in the development of an
EFT description of nucleon-nucleon (NN) scattering.  Two quite
different power-countings were developed for these processes,
Weinberg~\cite{Weinberg:rz,Weinberg:um,Ordonez:1992xp,Ordonez:tn} and
KSW
power-counting~\cite{Kaplan:1996xu,Kaplan:1998tg,Kaplan:1998we,Kaplan:1998sz},
with neither providing a complete description of scattering in all
partial waves.  A third power-counting, BBSvK
counting~\cite{Beane:2001bc}, which is essentially a synthesis of
these countings appears to be both formally consistent and provides a
convergent expansion. The difficulty in establishing an EFT
description is due to the large scattering lengths in both the $\si$
and $\siii$ channels, or equivalently, the near threshold (un)bound
states in both channels. Such states cannot be generated at any order
in perturbation theory, and require the non-perturbative resummation
of at least one operator in the theory.  Physically speaking, such
states result from a fine-tuning between kinetic and potential energy,
and imply that the EFT is in the proximity of an unstable infrared
fixed point~\cite{Birse} (and thus implies that QCD has an infrared unstable
fixed point).

There is no reason to suppose that the hyperon-nucleon sector is near
an infrared fixed point. That is to say, while the bounds of
eq.~(\ref{eq:lambdaNexpert}) do not preclude unnaturally large scattering 
lengths, we
would expect the scattering lengths to be of natural size, and for there to
be no $\Lambda N$ bound states near threshold.  Therefore, one expects
the EFT that describes the interaction between hyperons and nucleons
to have a power-counting based on the engineering dimensions of the
operators that appear in the effective Lagrange density, and hence to
be precisely the power-counting of
Weinberg~\cite{Weinberg:rz,Weinberg:um}~\footnote{Weinberg's power-counting does not, in general,
{\it require} resummation of interactions.}.  
As a well-defined procedure
exists, due to L{\"u}scher~\cite{Luscher:1986pf,Luscher:1990ux}, for
extracting the low-energy elastic scattering parameters from
finite-volume lattice QCD calculations, we set-up the low-energy EFT
describing $\Lambda N$ scattering in Weinberg
power-counting\footnote{Note, however, that we power-count the
amplitude directly, rather than the potential.}.  Given the
L{\"u}scher framework, we use the EFT to compute the scattering length
and effective range for elastic $\Lambda N$ scattering to NLO in the
EFT expansion~\footnote{The three-body system with one $\Lambda$-hyperon has
been investigated in an EFT without pions~\cite{Hammer:2001ng}.}. 
These parameters will be sufficient to describe
$\Lambda N$ scattering for energies below that necessary for inelastic
processes $\Lambda N\rightarrow\Sigma N$ and $\Lambda
N\rightarrow\Lambda N \pi$. (See also Ref.~\cite{rob}.) Our results contain the leading
light-quark mass dependence of low-energy $\Lambda N$ scattering and
can be used to extrapolate from lattice-quark mass values to nature.

\subsection{$\Lambda N\rightarrow \Lambda N$ in QCD}

\noindent If one could perform a lattice QCD simulation at the
physical values of the quark masses, then, barring the advent of new experiments, there would be no need to
construct the EFT for the $\Lambda N\rightarrow \Lambda N$ S-matrix at
low energies.  However, near-future lattice simulations will be
partially-quenched with unphysically large values of the sea-quark
masses. Therefore, at least in the short term, the EFT construction
will be needed to extrapolate to QCD. 

In defining the nonrelativistic fields appropriate for
computing the strong interactions between the $\Lambda$ and the
nucleon we remove the nucleon classical trajectory from the nucleon
field, and the $\Lambda$ classical trajectory from the $\Lambda$
field~\footnote{ Usually, one removes only the classical trajectory
associated with one of the heavy-hadron fields.  However, we will work
with two-flavor $\chi$PT in which strange-baryon number is conserved
at each interaction, and hence the $\Lambda N$ residual mass term,
$\Delta_{\Lambda N}$ can be removed from the theory by a phase
redefinition, leaving only the $\Sigma\Lambda$ residual mass term.  }.
We can perform these redefinitions as we will work with an $SU(2)_L\otimes
SU(2)_R$ chirally-invariant theory, as opposed to its three-flavor
analog. The Lagrange density describing the free-field dynamics of
the nucleon, $\Lambda$ and $\Sigma$ is
\begin{eqnarray}
{\cal L} & = & N^\dagger \left( i\partial_0 + {\nabla^2\over 2 M_N}\right) N
\ +\ \Lambda^\dagger \left( i\partial_0 + {\nabla^2\over 2 M_\Lambda} \right) \Lambda
\ +\ \Sigma^\dagger\ \left( i\partial_0 + {\nabla^2\over 2 M_\Lambda} -
  \Delta_{\Sigma\Lambda}\right)\Sigma
\ \ \ ,
\label{eq:free}
\end{eqnarray}
where $\Lambda$ is an iso-singlet while the nucleon iso-doublet and the
$\Sigma$ iso-triplet are defined as
\begin{eqnarray}
N & = & \left(\matrix{p\cr n}\right)
\ \ ,\ \ 
\Sigma \ =\ \left(
\matrix{\Sigma^0/\sqrt{2} & \Sigma^+\cr \Sigma^- & -\Sigma^0/\sqrt{2} }\right)
\ \ .
\label{eq:NSigdef}
\end{eqnarray}
As we are interested only in $\Lambda N$ scattering up to NLO, we need
not consider the Weinberg-Tomazawa type interactions.
The complete kinetic
energy terms for the $N$ and $\Sigma$ have additional multi-pion
interactions, but as the $\Lambda$
is an isosinglet its chirally-invariant kinetic energy term
is given in eq.~(\ref{eq:free}). 
The leading-order (LO) interactions between the baryons
and the pions are given by (to linear order in the pion field)
\begin{eqnarray}
{\cal L} & = & 
{g_A\over f}\ N^\dagger\sigma\cdot\nabla M N
\ +\ 
{g_{\Lambda\Sigma}\over f}\
\left( \ 
\Lambda^\dagger 
{\rm Tr}\left[\ 
\sigma\cdot\nabla M \Sigma\ \right]\ +\ {\rm h.c.}\right) 
\nonumber\\
&&
\ +\ 
{g_{\Sigma\Sigma}\over f}\
{\rm Tr}\left[\ \Sigma^\dagger \left[ \sigma\cdot\nabla M\ ,\ \Sigma\ \right]\right]
\ \ ,
\label{eq:su2axials}
\end{eqnarray}
where $f\sim 132~{\rm MeV}$ is the pion decay constant, 
$g_A\sim 1.27$ is the nucleon axial coupling constant, 
$g_{\Lambda\Sigma}\sim 0.60$ is the $\Lambda\Sigma$ axial coupling constant,
while $g_{\Sigma\Sigma}$ is the $\Sigma\Sigma$ axial coupling that 
cannot be determined directly from experiment and which
we will discuss later. $M$ is the matrix of pions given by
\begin{eqnarray}
M & = & \left(\matrix{\pi^0/\sqrt{2} & \pi^+\cr \pi^- & -\pi^0/\sqrt{2}}\right)
\ \ \ .
\label{eq:pions}
\end{eqnarray}

At LO in the expansion, there will be contributions to $S-$wave scattering from four-baryon
operators of the form
\begin{eqnarray}
{\cal L} & = & 
_{\Lambda\Lambda}C_0^{(\si)} 
\left( \ \Lambda^T\ P^{(\si)}\  N \right)^\dagger
\left( \ \Lambda^T \ P^{(\si)}\  N \right)
 +\ 
_{\Lambda\Lambda}C_0^{(\siii)} 
\left( \ \Lambda^T\ P^{(\siii)}\   N \right)^\dagger
\left( \ \Lambda^T\ P^{(\siii)}\   N \right)
\ \ \
\label{eq:lamlamints}
\end{eqnarray}
where 
$P^{(\si)}={1\over\sqrt{2}} \ i\sigma_2$ and 
$P^{(\siii)}={1\over\sqrt{2}} \ i\sigma_2\sigma^a $
are spin-projectors for the $\si$ and $\siii$ channels respectively,
and the coefficients $_{\Lambda\Lambda}C_0^{(\si,\siii)}$ are to be determined.
At next-to-leading order (NLO) there are also contributions from 
four-baryon operators that involve a $\Sigma$ baryon,
\begin{eqnarray}
{\cal L} & = & 
_{\Sigma\Lambda}C_0^{(\si)} 
\left(\ \Sigma^T(i\tau_2)\ P^{(\si)}\   N\right)^\dagger 
\left(\ \Lambda^T(i\tau_2)\ P^{(\si)}\  N\right)
\nonumber\\
& + &
_{\Sigma\Lambda}C_0^{(\siii)} 
\left(\ \Sigma^T(i\tau_2)\ P^{(\siii)}\  N\right)^\dagger 
\left(\ \Lambda^T(i\tau_2)\ P^{(\siii)}\   N\right)
\ +\ {\rm h.c.}
\ \ ,
\label{eq:siglamints}
\end{eqnarray}
where the coefficients $_{\Sigma\Lambda}C_0^{(\si,\siii)}$ are to be
determined. 

We have recognized and implemented a particular hierarchy of mass-scales in order to
simplify our expressions for the scattering length and effective range.
By returning briefly to $SU(3)$, we recall that the $\Sigma\Lambda$ mass
splitting, $\Delta_{\Lambda\Sigma}\sim m_q$ is proportional to the light-quark
masses, in particular, the strange quark mass, and numerically is 
$\Delta_{\Lambda\Sigma}\sim 74~{\rm  MeV}$.
In contrast, the pion mass is $m_\pi\sim m_q^{1/2}$ and numerically is 
$m_\pi\sim 140~{\rm  MeV}$.  Therefore, in what follows we will neglect 
$\Delta_{\Lambda\Sigma}$ compared to $m_\pi$.
However, we will retain terms of the form
$\sqrt{ 2 \mu_{\Lambda N}\Delta_{\Lambda\Sigma}}$ 
(where $\mu_{\Lambda N}$ is the reduced mass of the $\Lambda N$ system)
compared to $m_\pi$, as they
are of the same order in the $SU(3)$ chiral expansion.
While we can (somewhat) justify this expansion from a formal point of view, 
one should keep in mind that the expansion is in factors of $\sim 2$.
Of course the sole purpose of this expansion is to provide simple closed-form
expressions for the effective-range parameters; the exact numerical expressions are easily obtained and may in fact
be necessary in certain regions of the $C_0$ parameter space.

\begin{figure}[!ht]
\centerline{{\epsfxsize=1.15in \epsfbox{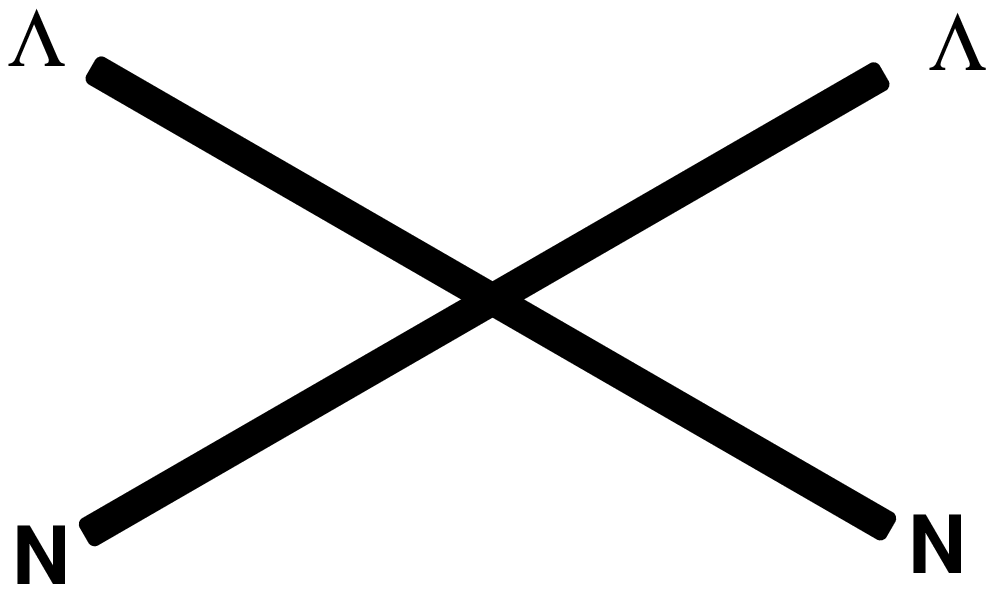}}} 
\vskip 0.2in
\centerline{{\epsfxsize=2.3in \epsfbox{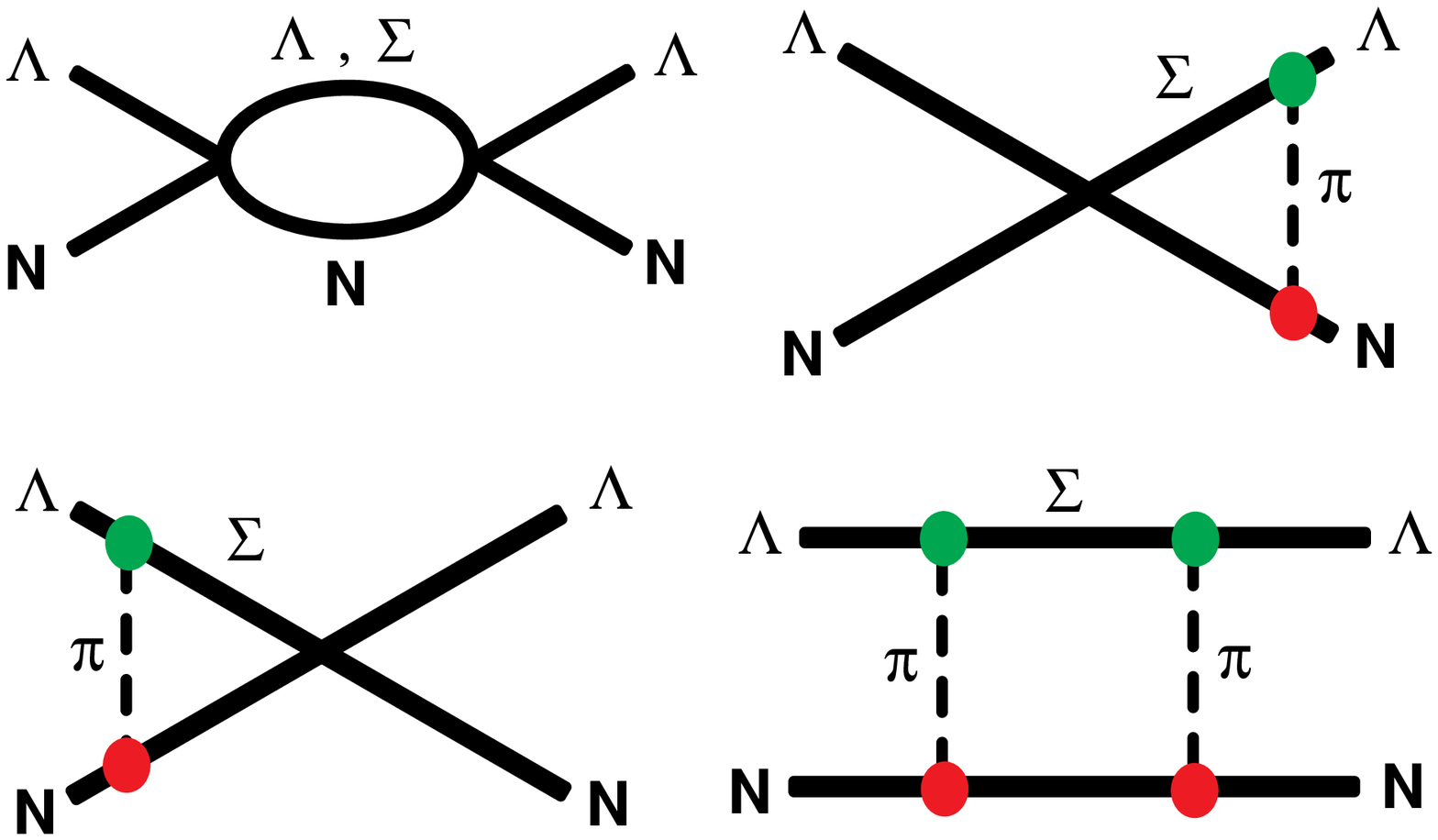}}} 
\vskip 0.15in
\noindent
\caption{\it 
Diagrams that contribute to $\Lambda N$ scattering at LO (top) and NLO in the EFT.
The shaded blobs are vertices from eq.~(\ref{eq:su2axials}).
}
\label{fig:LNLN}
\vskip .2in
\end{figure}

A straightforward calculation of the diagrams in Fig.~\ref{fig:LNLN} 
yields the scattering length $a^{(\si)}$ and effective range $r^{(\si)}$ in the $\si$ channel:
\begin{eqnarray}
a^{(\si)} & = & 
- {\mu_{\Lambda N}\over 2\pi}
\left[\ _{\Lambda\Lambda}C_0^{(\si)}
\ -\ 
{3\over 4\pi} \left({_{\Sigma\Lambda}C_0^{(\si)}}\right)^2\ \mu_{\Lambda N} \ \eta
\ +\ 
 {_{\Sigma\Lambda}C_0^{(\si)}}\  
{ 3 g_{\Sigma\Lambda}\  g_A \   \mu_{\Lambda N}\over 2 \pi f^2}\ 
{ \eta^2 + \eta m_\pi + m_\pi^2\over \eta+m_\pi}
\right.\nonumber\\ &&\left.
\ -\ 
{3 g_{\Sigma\Lambda}^2\  g_A^2\  \mu_{\Lambda N}\over 4 \pi f^4}\ 
{2\eta^3+ 4 \eta^2 m_\pi + 6 \eta m_\pi^2 + 3 m_\pi^3\over 2(\eta+m_\pi)^2}\ 
\right] \ ;
\nonumber\\
r^{(\si)} & = & 
- {1\over \mu_{\Lambda N}\ \pi}
\left[{2\pi\over  _{\Lambda\Lambda}C_0^{(\si)}}\right]^2
\left[\ 
{3\over 8\pi} \left({_{\Sigma\Lambda}C_0^{(\si)}}\right)^2\ {\mu_{\Lambda N}\over \eta} 
\ -\ 
 {_{\Sigma\Lambda}C_0^{(\si)}}\  
{ 3 g_{\Sigma\Lambda}\  g_A \   \mu_{\Lambda N}\over 2 \pi f^2}\ 
{3 \eta^2+ 9 \eta m_\pi + 8 m_\pi^2\over 6 (\eta+m_\pi)^3}
\right.\nonumber\\ &&\left.
\ +\ 
{3 g_{\Sigma\Lambda}^2\  g_A^2\  \mu_{\Lambda N}\over 4 \pi f^4}\ 
{6 \eta^3 + 23 \eta^2 m_\pi + 28 \eta m_\pi^2 + 7 m_\pi^3\over 12 (\eta+m_\pi)^4}
\ \right]
\ \ \ ,
\label{eq:1s0rangepars}
\end{eqnarray}
where $\eta = \sqrt{ 2 \mu_{\Lambda N}\Delta_{\Lambda\Sigma}}$.
We see that there are two unknown constants, 
$_{\Lambda\Lambda}C_0^{(\si)}$ and $_{\Sigma\Lambda}C_0^{(\si)}$, 
and two observables, 
$a^{(\si)}$ and $r^{(\si)}$.
In the spin-triplet channel we find
\begin{eqnarray}
a^{(\siii)} & = & 
- {\mu_{\Lambda N}\over 2\pi}
\left[\ _{\Lambda\Lambda}C_0^{(\siii)}
\ -\ 
{3\over 4\pi} \left({_{\Sigma\Lambda}C_0^{(\siii)}}\right)^2\ \mu_{\Lambda N} \ \eta
\ -\ 
 {_{\Sigma\Lambda}C_0^{(\siii)}}\  
{ g_{\Sigma\Lambda}\  g_A \   \mu_{\Lambda N}\over 2 \pi f^2}\ 
{ \eta^2 + \eta m_\pi + m_\pi^2\over \eta+m_\pi}
\right.\nonumber\\ &&\left.
\ -\ 
{3 g_{\Sigma\Lambda}^2 \ g_A^2 \ \mu_{\Lambda N}\over 4 \pi f^4}\ 
{2\eta^3+ 4 \eta^2 m_\pi + 6 \eta m_\pi^2 + 3 m_\pi^3\over 2(\eta+m_\pi)^2}\ 
\right]
\ \ ,
\end{eqnarray}
and
\begin{eqnarray}
r^{(\siii)} & = & 
- {1\over \mu_{\Lambda N} \pi}
\left[{2\pi\over  _{\Lambda\Lambda}C_0^{(\siii)} }\right]^2
\left[\ 
{3\over 8\pi} \left({_{\Sigma\Lambda}C_0^{(\siii)}}\right)^2\ {\mu_{\Lambda N}\over \eta} 
\ +\ 
 {_{\Sigma\Lambda}C_0^{(\siii)}}\  
{ g_{\Sigma\Lambda}\  g_A \   \mu_{\Lambda N}\over 2 \pi f^2}\ 
{3 \eta^2+ 9 \eta m_\pi + 8 m_\pi^2\over 6 (\eta+m_\pi)^3}
\right.\nonumber\\ &&\left.
\ -\ 
{3 g_{\Sigma\Lambda}^2\  g_A^2\  \mu_{\Lambda N}\over 4 \pi f^4}\ 
{46 \eta^3 + 91 \eta^2 m_\pi + 44 \eta m_\pi^2 + 11 m_\pi^3\over 36 (\eta+m_\pi)^4}
\ \right]
\ \  ,
\end{eqnarray}
where we again see that there are two unknown constants, 
$_{\Lambda\Lambda}C_0^{(\siii)}$ and $_{\Sigma\Lambda}C_0^{(\siii)}$, 
and two observables, $a^{(\siii)}$ and $r^{(\siii)}$.

It is worth commenting on the viability of the power-counting scheme
that we have chosen for the $\Lambda N$ system.  Consider the $\si$
effective-range parameters of eq.~(\ref{eq:1s0rangepars}). If we vary
the $C_0^{(\si)}$ coefficients over natural values (i.e. $\sim
1/f^2$), then there are large regions of parameter space where there
is poor convergence in the scattering length, and the effective range
is anomalously small and negative.  This may indicate that a
next-to-NLO (NNLO) calculation is required, which includes, {\it inter alia},
contact operators with two derivatives (the $C_2$ operators) and
contact operators with an insertion of the quark-mass matrix (the
$D_2$ operators). However, it may also be the case that we should apply our
power-counting scheme to a potential, rather than the amplitude
directly, and resum using the Schr{\"o}dinger equation, as in Weinberg
power-counting in the NN system.  The appropriate power-counting will
ultimately have to be decided in conjunction with lattice QCD simulations of
the $\Lambda N$ system.

\subsection{$\Lambda N\rightarrow \Lambda N$ in PQQCD}

\noindent While one always hopes for fully-unquenched lattice
simulations at the physical values of the light-quark masses, it will
always be the case that partially-quenched simulations of the
observables of interest will be performed first.  This is for the
simple reason that the partially-quenched simulations, in which the
masses of the sea-quarks are larger than those of the valence quarks,
take less time than the fully-unquenched simulations.  During the past
year or so a significant number of single-nucleon observables have
been explored in partially-quenched chiral perturbation theory
(PQ$\chi$PT)~\cite{Chen:2001yi,Beane:2002vq,Beane:2002ca} with an eye
to making connections between partially-quenched lattice simulations
and nature. The EFT describing NN interactions has also been partially
quenched~\cite{Beane:2002nu,Beane:2002np}.

\begin{figure}[!ht]
\centerline{{\epsfxsize=2.3in \epsfbox{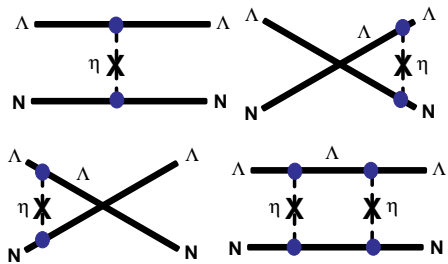}}} 
\vskip 0.15in
\noindent
\caption{\it 
The LO and NLO contributions to  $\Lambda N$ scattering from hairpin-exchange
in PQ$\chi$PT. The shaded blobs are vertices from eq.~(\ref{eq:etacouplings}).
}
\label{fig:HPbox}
\vskip .2in
\end{figure}

The partial quenching of the $\Lambda N$ scattering amplitude has
features from both the partially-quenched $NN$ scattering
amplitude~\cite{Beane:2002np} and from the $\Lambda_Q\Lambda_Q$
potential~\cite{Arndt:2003vx}.  The local four-baryon operators in 
PQQCD are the same as those in QCD because ghost-quark and sea-quark
number are conserved. It is for this reason that the pion box-diagram
is the same in PQQCD and QCD. The Crossed-Box diagram is
modified by partial quenching, but this effect enters at NNLO, one order
beyond the order to which we are working. Unlike the $\Lambda_Q\Lambda_Q$
potential~\cite{Arndt:2003vx}, there is a contribution from one-hairpin exchange (OHPE) at 
LO and also from one- and two-hairpin exchanges at NLO, as shown in Fig.~\ref{fig:HPbox}.

The $SU(2)_L\otimes SU(2)_R$ chiral symmetry of two-flavor QCD is
extended to the graded $SU(4|2)_L\otimes SU(4|2)_R$ chiral symmetry of
PQQCD, as discussed extensively in
Refs.~\cite{Beane:2002vq,Beane:2002ca}.  In this theory, in addition
to the light pions, light ghost-mesons and sea-mesons that contribute
in diagrams with single-pole propagators, there are also contributions
from $\eta$-exchange with a double-pole propagator of the form
\begin{eqnarray}
G_{\eta\eta} (q^2) & = & {i (m_{ss}^2 - m_\pi^2)\over (q^2-
  m_\pi^2+i\epsilon)^2}
 \ \ ,
\label{eq:etaprop}
\end{eqnarray}
where $m_{ss}$ is the mass of a meson composed only of sea quarks.
We are working in the isospin limit where the sea-quarks are
degenerate and the valence quarks are degenerate. This double-pole
arises because the singlet of $SU(2)$ is not the singlet of
$SU(4|2)$~\cite{Sharpe:2001fh}.  At LO in the chiral expansion, the
interaction between the $\eta$ and the baryons is described by a
Lagrange density of the form
\begin{eqnarray}
{\cal L} & = & 
{1\over \sqrt{2} f}\  g_0^{(N)}\   N^\dagger \sigma^k  N \  \nabla^k\eta
\ +\ 
{1\over \sqrt{2} f} g_0^{(\Lambda)}\   \Lambda^\dagger \sigma^k \Lambda
\nabla^k \eta
 \ \ ,
\label{eq:etacouplings}
\end{eqnarray}
where the axial couplings, $g_0^{(N,\Lambda)}$ are to be determined from
lattice simulations. The analogous contributions to the $\Lambda_Q\Lambda_Q$
potential~\cite{Arndt:2003vx}
are absent because the axial coupling of the hairpin to the $\Lambda_Q$
vanishes due to heavy-quark symmetry, and is non-zero only at order $1/m_Q$~\cite{Cho:1992cf}.
However, we expect the coupling $g_0^{(\Lambda)}$ to be significantly
smaller than $g_{\Sigma\Lambda}$ based on large-$N_C$ arguments.
In the large-$N_C$ limit, the ``size'' of the matrix
element of an operator with isospin $I$ and spin $J$ is of order 
${1/N_C^{|I-J|} }$~\cite{Aneesh,Kaplan:1995yg}. 
Hence we expect this axial coupling,
which has $I=0$ and $J=1$, to be suppressed by a factor of $\sim 3$ compared to
$g_{\Sigma\Lambda}$, which gives it a natural size of 
$g_0^{(\Lambda)}\sim 0.2$.

We find that the leading contributions to the scattering length and effective range in
the $\si$ channel due to partial quenching are 
\begin{eqnarray}
\delta a^{(\si)} & = & 
- {\mu_{\Lambda N}\over 2\pi}
\left[\ _{\Lambda\Lambda}C_0^{(\si)}
{g_0^{(\Lambda)} g_0^{(N)} \mu_{\Lambda N}\over 4\pi f^2}
{m_{ss}^2-m_\pi^2\over m_\pi}
\ +\ 
{(g_0^{(\Lambda)})^2 (g_0^{(N)})^2 \mu_{\Lambda N}\over  128\pi f^4}
{(m_{ss}^2-m_\pi^2)^2\over m_\pi^3}
\right] \ ;
\nonumber\\
\delta r^{(\si)} & = & 
- {1\over \mu_{\Lambda N} \pi}
\left[{2\pi\over  _{\Lambda\Lambda}C_0^{(\si)} }\right]^2
{g_0^{(\Lambda)} g_0^{(N)}\over f^2}
{m_{ss}^2-m_\pi^2\over m_\pi^4}
\ \ \ ,
\end{eqnarray}
and in the $\siii$ channel are
\begin{eqnarray}
\delta a^{(\siii)} & = & 
- {\mu_{\Lambda N}\over 2\pi}
\left[\ - _{\Lambda\Lambda}C_0^{(\siii)}
{g_0^{(\Lambda)} g_0^{(N)} \mu_{\Lambda N}\over 12\pi f^2}
{m_{ss}^2-m_\pi^2\over m_\pi}
\ +\ 
{(g_0^{(\Lambda)})^2 (g_0^{(N)})^2 \mu_{\Lambda N}\over  128\pi f^4}
{(m_{ss}^2-m_\pi^2)^2\over m_\pi^3}
\right] \ ;
\nonumber\\
\delta r^{(\siii)} & = & 
+ {1\over \mu_{\Lambda N} \pi}
\left[{2\pi\over  _{\Lambda\Lambda}C_0^{(\siii)} }\right]^2
{g_0^{(\Lambda)} g_0^{(N)}\over 3 f^2}
{m_{ss}^2-m_\pi^2\over m_\pi^4}
\ \ \ .
\end{eqnarray}
It is important to note that the additional contributions to the
scattering lengths arise from the NLO diagrams in
Fig.~\ref{fig:HPbox}, and as required, there is no contribution from
OHPE (due to the derivative coupling). By contrast, the contribution
to the effective ranges begins at LO from OHPE. Therefore,
partial-quenching will perturbatively modify the scattering lengths,
but overwhelmingly dominate the effective ranges, except very near the
QCD limit.  Given that extractions of both the scattering length and
effective range are required to determine the four-baryon,
momentum-independent leading-order interactions, partial-quenching may
provide a serious obstacle to this procedure.

\subsection{$\Lambda N\rightarrow \Lambda N$ at Finite Volume}

\noindent The Maiani-Testa theorem~\cite{Maiani:ca} precludes
determination of scattering amplitudes from Euclidean-space Green functions
at infinite volume away from kinematic thresholds.  While appearing to
be a significant setback for lattice QCD for which only Euclidean-space 
Green functions can be computed, it was realized by
L{\"u}scher~\cite{Luscher:1986pf,Luscher:1990ux} that one can access
$2\rightarrow 2$ scattering amplitudes from lattice calculations
performed at finite volume.  L{\"u}scher's work generalizes a result from
nonrelativistic quantum mechanics~\cite{yang} to quantum field theory.  The
results are obtained by assuming that the lattice spacing,
$b$~\footnote{In this work we use $b$ instead of $a$ to denote the
lattice spacing. Clearly this will be considered a sacrilege in the lattice
community, but denoting the lattice spacing by the traditional $a$
would lead to confusion with the scattering lengths which are also
traditionally denoted by $a$.}, is very small and the lattice size,
$L$, is much larger than the range of the potential between the two
particles.

During the past few years, the pionless effective field theory,
$\nopi$~\cite{Chen:1999tn,vanKolck:1998bw}, description of NN
scattering at momenta much less than the pion mass has been studied
extensively, and it can be applied to describe the finite-volume
construction described above~\footnote{This discussion will be the
same as that for pseudo-potentials~\cite{vanBaal:2000zc} as the
``physics'' is the same. However the $\nopi$ construction allows for
generalization to processes including gauge fields.}.  In $\nopi$ the
interactions between nucleons are described by a series of
delta-functions and their derivatives.  While the two-nucleon sector
reproduces identically effective-range theory, when gauge fields are
included one has a systematic and calculable scheme with which to
determine all low-energy multi-nucleon processes.  This theory has
been successfully applied to quite a wide variety of two-nucleon and
three-nucleon processes\footnote{One may worry about potential violations of unitarity in 
$\nopi$ for PQQCD. However, as pointed out in Ref.~\cite{Beane:2002np}, one can use
the methods of Ref.~\cite{Beane:2002nu} to show that all effects of
partial-quenching are in the coefficients of the contact operators. Therefore,
perhaps somewhat surprisingly, $\nopi$ is unitary in PQQCD.}. 
In order to establish a power-counting at
the level of diagrams in systems with unnaturally-large scattering
lengths, the power-divergence subtraction scheme (PDS) was introduced
to deal with the power-law ultra-violet divergences that appear in the
non-relativistic theory at the one-loop level.  It is somewhat useful
to see how one recovers L{\"u}scher's relations between the
energy-levels of the quantum system in a periodic box and the
scattering parameters of the continuum theory in $\nopi$, and this is
presented in Appendix I.  As re-derived in Appendix I, the energy of
the lowest-lying scattering state of the two-particle system with zero center of mass
momentum in a periodic box of sides $L$ that is much greater than any
of the scattering parameters, such as the scattering length and
effective range, is~\cite{Luscher:1986pf,Luscher:1990ux}
\begin{eqnarray}
E_0 & = & + {2\pi a\over \mu_{\Lambda N} \ L^3}\left[\ 1\ -\ c_1 {a\over L}\ 
+\ c_2 \left({a\over L}\right)^2\ +\ ...\right]
\ +\ {\cal O}(L^{-6})
\ \ \ ,
\label{eq:e0t}
\end{eqnarray}
where the coefficients $c_{1,2}$ are 
$c_1=-2.837297$ and $c_2=6.375183$,
and where we have used the conventional definition of scattering length and
effective range 
\begin{eqnarray}
p\cot\delta & = & -{1\over a} + {1\over 2} r p^2 + ...
\ \ \ .
\label{eq:er}
\end{eqnarray}
The energy of the next level which transforms in the $A_1$ representation of the
cubic group is~\cite{Luscher:1986pf,Luscher:1990ux}
\begin{eqnarray}
E_1 & = & {2\pi^2\over \mu_{\Lambda N} \ L^2} - 
{6\tan\delta_0\over \mu_{\Lambda N} \ L^2}\left(\ 
1 + c_1^\prime\tan\delta_0 + c_2^\prime \tan^2\delta_0\ +\ ...\ \right)
\ +\ {\cal O}(L^{-6})
\ \ \ ,
\label{eq:e1t}
\end{eqnarray}
where $c_1^\prime = -0.061367$ and $c_2^\prime = -0.354156$,
and where $\delta_0$ is the $S-$wave phase shift evaluated at the unperturbed
lattice momentum $|{\bf p}| = 2\pi/L$. It is important to stress that the
effective-range parameters may be comparable to or greater than realistic box sizes at some
values of the unphysical lattice quark masses. Eqs.~(\ref{eq:e0t}) and (\ref{eq:e1t}) 
are then no longer applicable and one must use the exact formula for the energy levels, eq.~(\ref{eq:energies}), given
in Appendix I.

\section{Nonleptonic Weak Decays and Weak Hypernuclear Decays}

\noindent The weak decays of hypernuclei result not only from the
one-body nonleptonic decays (mesonic) of the hyperon(s),
e.g. $\Lambda\rightarrow N\pi$, that exists in free-space but also
from two- and higher-body decays (non-mesonic), such as $\Lambda
N\rightarrow NN$ and $\Lambda NN\rightarrow NNN$.  Naive expectations
suggest that the one-body decays should dominate, however the
non-mesonic processes are found to be comparable to the mesonic decays
in nuclei, in part due to Pauli-blocking of the final-state nucleon in
mesonic decays.

The nonleptonic weak decays of hyperons in free-space have been
very-well studied during the past many decades.  While the $S-$ and
$P-$wave amplitudes that contribute to each process are very-well
known experimentally, the $P-$wave amplitudes are still not well
described theoretically.  In order for one to claim that the weak
decays of hypernuclei are theoretically understood we must first
understand the nonleptonic weak decays of hyperons in free-space, as
this underlies one of the two competing decay processes.  Furthermore,
the EFT that describes these processes is more than likely entangled
in the theory that describes the higher-body amplitudes, and therefore
is central to the description of all decay processes.

\subsection{Nonleptonic Weak Decays of Hyperons in Free-Space}

\noindent As the nonleptonic decays $\Lambda\rightarrow p\pi^-$,
$\Sigma^-\rightarrow n\pi^-$, $\Sigma^+\rightarrow n\pi^+$,
$\Xi^-\rightarrow\Lambda\pi^-$ and their isospin partners have been
well studied for an extensive period of time, the experimental data
and theoretical efforts to understand the data can be found in many
textbooks, e.g. Ref.~\cite{Donoghue:dd}.  Therefore, we do not attempt
an exhaustive discussion of the existing literature in this work.
Angular-momentum considerations are sufficient to write the matrix
element for the nonleptonic decay $B\rightarrow B^\prime\pi$, ${\cal
M}_{BB^\prime\pi}$, using the notation of
Jenkins~\cite{Jenkins:1991bt}, as
\begin{eqnarray}
{\cal M} & = & G_F m_{\pi^+}^2 \ 
\overline{U}_{B^\prime}\ 
\left[
{\cal A}^{(S)}\  + \ 2\  S^\mu \hat k_\mu \ {\cal A}^{(P)}\ \right]\ 
U_B
\ \ \ ,
\label{eq:weakmat}
\end{eqnarray}
where $\hat k$ is the four-vector of the outgoing
pion normalized to the pion three-momentum in the baryon rest frame.
The factor of $m_{\pi^+}^2$ that appears in the coefficient in
eq.~(\ref{eq:weakmat}) is the physical mass of the charged pion, and
does not indicate that the amplitude vanishes in the chiral limit.
The amplitudes ${\cal A}^{(S)}$ and ${\cal A}^{(P)}$ are the $S-$ and
$P-$wave amplitudes, respectively.  They are expected to be the same
order in the chiral expansion, and indeed experimentally, their
magnitudes are of order unity.

\subsubsection{Nonleptonic Weak Decays of Hyperons in $SU(3)$}

\begin{figure}[!ht]
\centerline{{\epsfxsize=1.15in \epsfbox{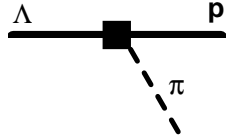}}} 
\vskip 0.15in
\noindent
\caption{\it 
Tree-level diagram that gives the leading-order contribution to the $S-$wave
amplitudes for the nonleptonic decay $\Lambda\rightarrow p\pi^-$. 
The solid square denotes an insertion of the leading-order weak interaction
from eq.~(\ref{eq:weakcouplings}).
}
\label{fig:swave3}
\vskip .2in
\end{figure}

\begin{figure}[!ht]
\centerline{{\epsfxsize=2.9in \epsfbox{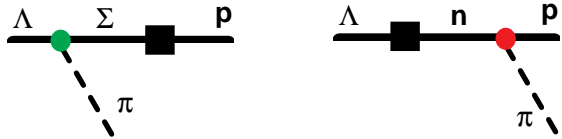}}} 
\vskip 0.15in
\noindent
\caption{\it 
Tree-level diagrams that give the leading-order contribution to the $P-$wave
amplitudes for the nonleptonic decay $\Lambda\rightarrow p\pi^-$. 
The solid-square denotes an insertion of the leading-order weak interaction
from eq.~(\ref{eq:weakcouplings}) while the shaded circles denote axial couplings.
}
\label{fig:pwave3}
\vskip .2in
\end{figure}

\noindent Conventionally, the theoretical analysis of these decays is
performed about the limit of $SU(3)$ flavor symmetry, assuming the
$\Delta I={1\over 2}$ rule, although a more general analysis shows
that the $\Delta I={1\over 2}$ rule is well respected in these
decays~\cite{Donoghue:dd}.  At LO in the chiral expansion, the weak
decays are described by the Lagrange density
\begin{eqnarray}
{\cal L} & = & 
G_F \ m_{\pi^+}^2 \ f\  
\left(\ h_D\  {\rm Tr}\left[\overline{B}\{h_\xi , B\}\right]
\ +\ 
h_F \ {\rm Tr}\left[\overline{B}\left[h_\xi , B\right]\right]
\ \right)
\ \ \ ,
\label{eq:weakcouplings}
\end{eqnarray}
where
\begin{eqnarray}
h_\xi & = &\xi^\dagger \ h \ \xi
\ \ \ \ {\rm with}\ \ \ \
h \ = \ \left(\matrix{0&0&0\cr 0&0&1\cr 0&0&0}\right)
\ \ \ ,
\end{eqnarray}
and where $h$ transforms as a $({\bf 8},{\bf 1})$, $h\rightarrow
LhL^\dagger$, under $SU(3)_L\otimes SU(3)_R$ chiral transformations.
Expanding this interaction to linear order in the pion field, it
is straightforward to show that the $S-$amplitudes arising from the
diagram in Fig.~\ref{fig:swave3} or its analog
are~\cite{Jenkins:1991bt,Jenkins:1991ne}
\begin{eqnarray}
\begin{array}{ll}
{\cal A}^{(S)}_{\Lambda p\pi^-}
 \ =\   -{1\over \sqrt{6}} \left( h_D+3h_F\right)
\ \  & 
{\cal A}^{(S)}_{\Sigma^-n\pi^-}
 \ =\  h_D-h_F
\\
{\cal A}^{(S)}_{\Sigma^+n\pi^+}
\ =\   0
&{\cal A}^{(S)}_{\Xi^-\Lambda\pi^-}
\ = \ {1\over \sqrt{6}} \left( 3h_F-h_D\right)
\end{array}
\end{eqnarray}
and the $P-$wave amplitudes arising from the diagrams in Fig.~\ref{fig:pwave3}
and their analogs
are 
\begin{eqnarray}
{\cal A}^{(P)}_{\Lambda p\pi^-}
& = & 
{\sqrt{2\over 3}D (h_D-h_F)\over M_p-M_{\Sigma^+}}
 \ -  \ 
{(D+F)(h_D+3h_F)\over\sqrt{6} (M_\Lambda-M_n)}
\nonumber\\
{\cal A}^{(P)}_{\Sigma^- n\pi^-}
& = & 
{ F (h_F-h_D)\over M_n-M_{\Sigma^0}} - 
{D(h_D+3h_F)\over 3(M_n-M_\Lambda)}
\nonumber\\
{\cal A}^{(P)}_{\Sigma^+ n\pi^+}
& = & 
-{D(h_D+3h_F)\over 3(M_n-M_\Lambda)}
-{ F (h_F-h_D)\over M_n-M_{\Sigma^0}}
+{(D+F) (h_D-h_F)\over M_{\Sigma^+} - M_p}
\nonumber\\
{\cal A}^{(S)}_{\Xi^-\rightarrow \Lambda\pi^-}
& = & 
{ (D-F)(3 h_F-h_D)\over\sqrt{6}(M_\Lambda-M_{\Xi^0})}
+ \sqrt{2\over 3}{ D (h_D+h_F)\over M_{\Xi^-}-M_{\Sigma^-}}
\ \ \ ,
\end{eqnarray}
where $D$ and $F$ describe the matrix elements of the octet axial current
operators between octet baryons.

In table~\ref{table:su3} we present the numerical values for the 
LO tree-level amplitudes, along with their experimental values.
We choose to determine $h_D$ and $h_F$ from the $S-$wave amplitudes, which give central values
$h_D=0.58$ and $h_F=-1.40$, and to use the tree-level values for the axial
matrix elements, $D= 0.80$ and $F=0.50$~\footnote{
Ref.~\cite{Savage:1996zd} finds $D=0.79\pm 0.10$ and $F=0.47\pm
0.07$ from a  global fit of tree-level axial matrix elements to the available 
semileptonic decay data.}.
\begin{table}[ht]
\caption{Weak Amplitudes in $SU(3)$ $\chi$PT at LO and NLO}
\label{table:su3}
\newcommand{\m}{\hphantom{$-$}}
\newcommand{\cc}[1]{\multicolumn{1}{c}{#1}}
\renewcommand{\tabcolsep}{0.2pc} 
\renewcommand{\arraystretch}{1.0} 
\begin{tabular}{@{}c | c | c | c | c | c | c }
\hline
Decay & 
${\cal A}^{(S)}$ LO & 
${\cal A}^{(S)}$ NLO~\cite{Springer:sv} & 
${\cal A}^{(S)}$ Expt & 
${\cal A}^{(P)}$ LO & 
${\cal A}^{(P)}$ NLO~\cite{Springer:sv} & 
${\cal A}^{(P)}$ Expt\\
\hline
$\Lambda\rightarrow p\pi^-$ & 
1.48 & 1.44 & $1.42\pm 0.01$ & 
0.59 & $-0.73\pm 0.18$  & $0.52\pm 0.02$ \\
$\Sigma^-\rightarrow n\pi^-$ & 
1.98 & 1.89 & $1.88\pm 0.01$ & 
-0.30 & $0.46\pm 0.21$ & $-0.06\pm 0.01$ \\
$\Sigma^+\rightarrow n\pi^+$ & 
0.0 & 0.01 & $0.06\pm 0.01$ & 
0.16 & $-0.18\pm 0.21$ & $1.81\pm 0.01$ \\
$\Xi^-\rightarrow \Lambda \pi^-$ & 
-1.95 & -2.01 & $-1.98\pm 0.01$ & 
-0.19 & $0.52\pm 0.29$ & $0.48\pm 0.02$ \\
\hline
\end{tabular}
\end{table}
Also, in table~\ref{table:su3} we present the amplitudes that are found at
one-loop order (NLO) in the three-flavor chiral expansion~\cite{Springer:sv} (see
also Ref.~\cite{AbdEl-Hady:1999mj}).

The numerical values in table~\ref{table:su3} make it clear that the
theoretical analysis does not come close to predicting the $P-$wave
amplitudes. Moreover, the one-loop computation in three-flavor
$\chi$PT~\cite{Jenkins:1991ne,Springer:1995nz,Springer:sv,AbdEl-Hady:1999mj}
does not improve the situation and, arguably, makes the situation
worse.  There are a variety of reasons one can conjecture as a
possible explanation of these discrepancies, e.g.  $SU(3)$ breaking is
abnormally large, there are strong final state interactions, there are
narrow hadronic resonances (e.g. pentaquarks) in some channels, and so
forth.  From a formal standpoint, it is possible that the three-flavor
$\chi$PT expansion simply does not behave well for these observables,
given that it is the kaon mass that is governing the convergence.
Jenkins~\cite{Jenkins:1991bt} has argued that convergence is poor, not
due to failure of the chiral expansion, but due to large cancellations
at LO.  In order to explore this possibility further, a NNLO
calculation is ultimately required.  As parameters that enter into the
NLO calculations were determined from a global fit it is somewhat
difficult to determine if the problems in the $P-$waves lie in only
one of the decay modes or if the theory fails equally badly in all
modes.

\subsubsection{Nonleptonic Weak Decays of Hyperons in $SU(2)$}

\noindent Given the absence of complete theoretical control in the
nonleptonic decays in three-flavor $\chi$PT, we de-scope our efforts
to two-flavor $\chi$PT with the hope of gaining some insight into
which processes can be identified as problematic and which are under
control, without having to consider the issue of $SU(3)$ breaking.  We
restrict ourselves to the decays of the $\Lambda$ and $\Sigma$ as we
are primarily interested in the study of hypernuclei containing these
hyperons.

At LO in the two-flavor chiral expansion, the weak Lagrange density is
\begin{eqnarray}
{\cal L} & = & 
G_F m_{\pi^+}^2 f 
\left[\ h_\Lambda\  \left(\overline{N}h_\xi\right)\Lambda
\ +\ 
h_\Sigma\  \overline{N}\Sigma h_\xi\ 
+\ {\rm h.c.}\ \right]
\ \ \ ,
\label{eq:su2weak}
\end{eqnarray}
where
\begin{eqnarray}
h_\xi & = & \xi^\dagger h
\ \ {\rm and}\ \ 
h\ =\ \left(\matrix{0\cr 1}\right)
\ \ \ ,
\end{eqnarray}
and where $h$, which transforms as 
$h\rightarrow L h$  under
$SU(2)_L\otimes SU(2)_R$ chiral transformations, is a $({\bf 2} , {\bf 1})$.  
As in the
three-flavor case, we have assumed a $\Delta I={1\over 2}$ weak
interaction only.

The weak interactions described in eq.~(\ref{eq:su2weak}) along with the strong
interactions between the pions and the baryons in eq.~(\ref{eq:su2axials}) 
give $S-$wave  amplitudes
\begin{eqnarray}
\begin{array}{lll}
{\cal A}^{(S)}_{\Lambda p\pi^-}
\ = \ h_\Lambda\ \ \ ,
&
{\cal A}^{(S)}_{\Sigma^- n\pi^-}
\ = \ h_\Sigma\ \ \ ,
&
{\cal A}^{(S)}_{\Sigma^+ n\pi^+}
\ =\ 0
\end{array}
\  \ \ ,
\label{eq:su2S}
\end{eqnarray}
and $P-$wave  amplitudes
\begin{eqnarray}
{\cal A}^{(P)}_{\Lambda p\pi^-}
& = & 
{g_A h_\Lambda\over M_\Lambda-M_n} 
+ {g_{\Lambda\Sigma} h_\Sigma\over M_p-M_{\Sigma^+}}
\ \ ,\ \ 
{\cal A}^{(P)}_{\Sigma^- n\pi^-}
\ = \ 
{g_{\Lambda\Sigma} h_\Lambda\over M_n-M_\Lambda}
- {g_{\Sigma\Sigma} h_\Sigma \over M_n-M_{\Sigma^0}} \ ,
\nonumber\\
{\cal A}^{(P)}_{\Sigma^+ n\pi^+}
& = & 
{g_{\Lambda\Sigma} h_\Lambda\over M_n-M_\Lambda}
+ { g_{\Sigma\Sigma} h_\Sigma \over M_n-M_{\Sigma^0}}
+ { g_A h_\Sigma\over M_{\Sigma^+} - M_p}
\ \ \ .
\label{eq:su2P}
\end{eqnarray}
We choose to fit the weak couplings $h_{\Lambda,\Sigma}$ to the central values
of the $S-$wave amplitudes, and find that $h_\Lambda=1.42$ and  $h_\Sigma=1.88$.

While the axial matrix elements $g_A=1.27$ and
$g_{\Lambda\Sigma}=0.60$ are well-known experimentally, only an upper
limit currently exists for $\Sigma^-\rightarrow\Sigma^0
e\overline{\nu}$, and thus there is no direct measurement of
$g_{\Sigma\Sigma}$.  Given the success of $SU(3)$ symmetry in relating
the axial-current matrix elements, it is worth exploring its
predictions for $g_{\Sigma\Sigma}$, along with the best estimates for
leading-order $SU(3)$ breaking.  In the limit of exact $SU(3)$, the
axial-coupling would be $g_{\Sigma\Sigma}=F$, and thus
$g_{\Sigma\Sigma}=0.5$ at leading order.  We focus on two quite
different approaches to estimating the $SU(3)$ breaking.  In one
method~\cite{Savage:1996zd}, the one-loop corrections to all the axial
matrix elements were computed in three-flavor $\chi$PT including the
decuplet as dynamical fields, and parameters in the one-loop
contributions were varied over a reasonable range in order to estimate
the size of $SU(3)$ breaking effects.  This analysis yields a range
for $g_{\Sigma\Sigma}$, $0.35\lsim g_{\Sigma\Sigma}\lsim 0.55$.  However, the
analysis is somewhat ad hoc and depends upon the convergence of the
three-flavor chiral expansion, but formally captures the leading-order
$SU(3)$ breaking in the chiral limit.  The other
method~\cite{Dai:1995zg,Flores-Mendieta:1998ii} we use to estimate
$g_{\Sigma\Sigma}$ is to perturb about the large$-N_C$ limit of QCD, a
program that has been very successful in understanding known baryon
properties, and has provided insight into the successes of the naive
constituent quark model.  This large$-N_C$ analysis of the axial
matrix elements leads to a range~\footnote{By extending the contents
of table II in Ref.~\cite{Flores-Mendieta:1998ii} the coupling
$g_{\Sigma\Sigma}$ is found to be
\begin{eqnarray}
g_{\Sigma\Sigma} & = &
{2\over 3} a + b + {4\over 3} c_3 - {2\over 3} c_4
\ \ \ .
\end{eqnarray}
Using the central values of the 
global fit values for the parameters $a, b, c_3$ and $c_4$ found in 
table VII of Ref.~\cite{Flores-Mendieta:1998ii} we arrive at the range for 
$g_{\Sigma\Sigma}$. We thank Aneesh Manohar and Elizabeth Jenkins for assisting
with this determination.}
$0.30\lsim g_{\Sigma\Sigma}\lsim 0.36$~\footnote{
For completeness we have also determined the axial coupling $g_\Xi$ that
appears in the interaction Lagrange density
\begin{eqnarray}
{\cal L}_\pi & = & 
{g_\Xi\over f}\  \overline{\Xi}\  \sigma\cdot\nabla M \ \Xi
\ \ ,
\end{eqnarray}
where $\Xi$ denotes the iso-doublet containing the $\Xi^0$ and $\Xi^-$.
By extending the contents of table II in 
Ref.~\cite{Flores-Mendieta:1998ii} the coupling $g_\Xi$ is found to
be
\begin{eqnarray}
g_{\Xi} & = & \left[\
{1\over 3} a - b + {4\over 3} c_3 - {8\over 3} c_4\ \right]
\ \ \ .
\end{eqnarray}
Using the central values of the 
global-fit values for the parameters $a, b, c_3$ and $c_4$ found in 
table VII of Ref.~\cite{Flores-Mendieta:1998ii} we find
$0.26\lsim g_\Xi\lsim 0.30$.
Further, the one-loop analysis of Ref.~\cite{Savage:1996zd} gives
$0.18\lsim g_\Xi\lsim 0.36$.
These two estimates are compatible with each other, with, as before, the 
estimate from Ref.~\cite{Savage:1996zd} having a larger range than that from 
Ref.~\cite{Flores-Mendieta:1998ii}.}.
Given that these distinct estimates of $g_{\Sigma\Sigma}$ have only a
small overlap and neither are obviously incorrect, we take $0.30\lsim
g_{\Sigma\Sigma}\lsim 0.55$ as the range of this axial coupling
constant, and it is clear that it is quite uncertain at present.

Inserting numerical values for the couplings and masses into the $S-$wave and
$P-$wave amplitudes in eq.~(\ref{eq:su2S}) and eq.~(\ref{eq:su2P}) gives the
numerical values for the nonleptonic amplitudes shown in table~\ref{table:su2}.
\begin{table}[ht]
\caption{Weak Amplitudes in $SU(2)$ $\chi$PT at LO ($g_{\Sigma\Sigma}=0.30\rightarrow 0.55$)}
\label{table:su2}
\newcommand{\m}{\hphantom{$-$}}
\newcommand{\cc}[1]{\multicolumn{1}{c}{#1}}
\renewcommand{\tabcolsep}{1pc} 
\renewcommand{\arraystretch}{1.2} 
\begin{tabular}{@{}c | c | c | c | c}
\hline
Decay & ${\cal A}^{(S)}$ Theory & ${\cal A}^{(S)}$ Expt & ${\cal A}^{(P)}$
Theory & 
${\cal A}^{(P)}$ Expt\\
\hline
$\Lambda\rightarrow p\pi^-$ & 1.42 (input) & $1.42\pm 0.01$ & 0.56 & $0.52\pm 0.02$ \\
$\Sigma^-\rightarrow n\pi^-$ & 1.88 (input) & $1.88\pm 0.01$ & 
$-0.50\rightarrow -0.14$ & $-0.06\pm 0.01$ \\
$\Sigma^+\rightarrow n\pi^+$ & 0.0  & $0.06\pm 0.01$ & 
$+0.42\rightarrow +0.08$  & $1.81\pm 0.01$ \\
\hline
\end{tabular}
\end{table}
At leading order, the $P-$wave amplitude in $\Lambda$ decays is well
predicted, as it is in the three-flavor theory.  However, we do not
expect significant modifications to this result from higher orders in
the two-flavor theory.  Further, we see that the $P-$wave amplitude
for $\Sigma^-\rightarrow n\pi^-$ is quite sensitive to the value of
$g_{\Sigma\Sigma}$, which is presently quite
uncertain.  At the upper limit of the allowed range for
$g_{\Sigma\Sigma}$, the $P-$wave amplitude is close to what is
observed.  Finally, we see that the $P-$wave amplitude for
$\Sigma^+\rightarrow n\pi^+$ is not close to the experimental
value for any reasonable value of $g_{\Sigma\Sigma}$, and theory
underestimates the experimental value by $\sim 4$ in the best case.

\subsection{Nonleptonic Weak Decays of Hyperons and  Lattice QCD}

\noindent Given the failure of both $SU(2)$ and $SU(3)$ $\chi$PT to
describe the $P-$wave amplitudes of the nonleptonic decays, one would
like to understand where the problems lie.  Until this issue is
resolved, one is unable to claim theoretical control in the nonleptonic
decays of hypernuclei.  Lattice QCD is, of course, the only rigorous
theoretical technique that exists to compute these decays, and it
is important to use lattice simulations as a diagnostic
of the problem(s) with the $\chi$PT calculations.

\begin{figure}[!ht]
\centerline{{\epsfxsize=2.9in \epsfbox{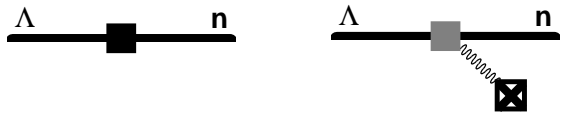}}} 
\vskip 0.15in
\noindent
\caption{\it 
Tree-level diagrams that contribute to the 
weak $\Lambda\rightarrow N$ matrix element, including the leading-order surface
term.
The dark shaded square denotes an insertion of the weak vertex, $h_\Lambda$
from eq.~(\ref{eq:su2weak}), while the light shaded square denotes an insertion of the weak vertices,
$j_\Lambda^{(1,2)}$, from eq.~(\ref{eq:su2surface}). The crossed-square denotes an injection of energy at the
weak vertex.
}
\label{fig:2pt}
\vskip .2in
\end{figure}

One should first compute the two-point functions that
arise from the leading-order interactions in eq.~(\ref{eq:su2weak}).
The lattice calculation of this process injects energy at
the weak four-quark operator in order to allow
both the initial and final states to be on mass-shell. Hence one must
consider additional operators in the EFT that are usually discarded as
surface terms.  In the $SU(2)$ expansion, the mass difference between
the $\Lambda$ or $\Sigma$ and the nucleon is taken to be a scale that
is fixed in the chiral limit, with dependence on the quark masses
arising at higher orders in the chiral expansion, i.e. the strange
quark mass is fixed. One therefore has the free Lagrange density
\begin{eqnarray}
{\cal L} & = & N^\dagger iv\cdot D N\ +\ \Lambda^\dagger i
v\cdot\partial\Lambda
\ -\ \Delta_{\Lambda N} \Lambda^\dagger\Lambda
\ +\ ...
\ \ \ .
\end{eqnarray}
At NLO there are contributions from
\begin{eqnarray}
{\cal L} & = & 
G_F m_{\pi^+}^2
\left[\ 
j_\Lambda^{(1)}\  
i v\cdot\partial\left[\ 
\left(\overline{N}h_\xi\right)\Lambda\ \right]
\ +\ 
j_\Lambda^{(2)}\  
\Delta_{\Lambda N}\left(\overline{N}h_\xi\right)\Lambda\ 
\right.\nonumber\\ &&\left.
\qquad \qquad 
\ +\ 
j_\Sigma^{(1)}\  
i v\cdot\partial\left[\ 
\overline{N}\Sigma h_\xi\ \right]
\ +\ 
j_\Sigma^{(2)}\  
\Delta_{\Sigma N}
\overline{N}\Sigma h_\xi\ 
+\ {\rm h.c.}\ \right]
\ \ \ ,
\label{eq:su2surface}
\end{eqnarray}
where the surface terms vanish
when energy and
momentum are conserved at the effective weak vertex.  However, in
lattice calculations, where both the hyperon and the nucleon will be
on mass-shell, these operators make a non-zero contribution to the
amplitude at tree-level.  This contribution must be included when
attempting to extract the weak couplings $h_{\Lambda,\Sigma}$ from
lattice calculations.  Therefore, including such terms, the weak
matrix element between a $\Lambda$ and a nucleon arising from the
diagrams shown in Fig.~\ref{fig:2pt}
becomes
\begin{eqnarray}
\langle  N\ ({\bf p}=0)  
|\  {\cal L}_{\rm weak}\  | \Lambda\ ({\bf p}=0) \rangle & = & 
G_F m_{\pi^+}^2 \left[\ f \ h_\Lambda\ +\ \Delta_{\Lambda N}\  
\left( j_\Lambda^{(1)}\ +\  j_\Lambda^{(2)}\ \right)
\right]
\ \overline{U}_N U_\Lambda .
\end{eqnarray}

Given that the $P-$wave amplitude for $\Lambda\rightarrow p\pi^-$ is
in good agreement with data, one expects that the value of $h_\Lambda$
extracted from the lattice will be in agreement with that determined
from the $S-$wave amplitudes.  However, it is not clear what one will
find for $h_\Sigma$ given the significant uncertainty in
$g_{\Sigma\Sigma}$ and the complete failure to reproduce the $P-$wave
amplitude for $\Sigma^+\rightarrow n\pi^+$.  Therefore, a lattice
determination of $h_\Sigma$ is highly desirable.  Furthermore, given that
there is a significant uncertainty in $g_{\Sigma\Sigma}$ and this
uncertainty propagates through into a large uncertainty in the
$P-$wave amplitudes for the $\Sigma$ decays, a lattice determination
of $g_{\Sigma\Sigma}$ is also highly desirable.

A direct lattice measurement of both the $S-$wave and $P-$wave
amplitudes would also be very important.  Assuming that the weak
couplings $h_{\Lambda,\Sigma}$ and the axial couplings are all
consistent with the above discussion, then we need to understand the
origin of the failure of the $P-$wave amplitudes.  Given the progress
in extracting weak amplitudes for $K\rightarrow\pi\pi$ using the
finite-volume techniques described by L{\"u}scher and
Lellouch~\cite{Lellouch:2000pv} one can hope for a direct lattice
determination of these amplitudes at some point in the future.  While
the actual value of the amplitude should agree with experiment 
one should be able to disentangle the various contributions to
the amplitudes and understand where the failure originates in the
low-energy EFT.

\subsection{Weak Decays of Hypernuclei}

\begin{figure}[!ht]
\centerline{{\epsfxsize=1.9in \epsfbox{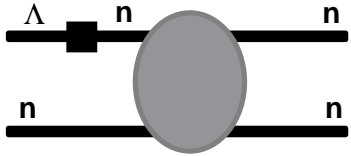}}} 
\vskip 0.15in
\noindent
\caption{\it 
An example of a contribution to $\Lambda n\rightarrow
nn$.
}
\label{fig:NNNLpole}
\vskip .2in
\end{figure}

\noindent A systematic study of the decays of hypernuclei provides an
even greater challenge than describing the nonleptonic decays of
hyperons in free-space.  It will certainly be the case that the
interactions that are ultimately needed to describe the nonleptonic
decays will also provide a contribution, and likely a leading-order
contribution, to the multi-baryon weak interactions such as $\Lambda
N\rightarrow NN$.  For instance, the diagram shown in
Fig.~\ref{fig:NNNLpole}
arises from inserting the weak two-point amplitude on an external leg
of the strong $nn\rightarrow nn$ scattering amplitude.  Unfortunately,
there is presently no consistent EFT that is applicable in the
kinematic regime for $\Lambda N\rightarrow NN$ where the energy
release for an initial $\Lambda N$ state at rest is $E\sim 176~{\rm
MeV}$. If an EFT exists for the strong processes, such as $\Lambda
N\rightarrow\Lambda N$ and $NN\rightarrow NN$ at these relatively high
energies, one hopes that the higher-dimension weak operators are
subleading compared to the two-point function, but given the
uncertainty in the nonleptonic hyperon decays and the absence of a
strong-interaction EFT, it appears premature to make any such
power-counting arguments at this stage.

However, having made all of these negative statements based on our
current theoretical understanding, there has been an important
phenomenological analysis~\cite{Parreno:2003ny,Parreno:2003mf}
recently completed by Parre\~no, Bennhold and Holstein (PBH) that
gives one hope that an EFT can in fact be constructed to describe these
processes.  In PBH, contributions to the weak decay from weak-one-pion
exchange (WOPE) and weak-one-kaon exchange (WOKE) were included along
with contributions from local $\Delta s=1$ weak $\Lambda NNN$
interactions.  The power-counting used in determining the order of the
various contributions is based on Weinberg's power-counting of
NN interactions in which the number of powers of the
small momenta is directly related to the order in the expansion.  The
weak interaction computed to a given order in Weinberg power-counting
is then dressed with NN strong interactions in the final state and the
initial-state interactions are taken into account by a correlated
$\Lambda N$ wavefunction determined from Nijmegen models.

The usual divergence problems with local four-baryon interactions
arise in this calculation and are dealt with in PBH by extending the
interaction in coordinate space to a Gaussian form with the same
spatial integral (of course, the dependence upon the form of the
regulator must be, and is, higher order in the power-counting).  The results
of a global fit, and data can be found in table~\ref{table:hyper}.
\begin{table}[ht]
\caption{Weak decay observables
for $^5 _\Lambda {\rm He}$,
$^{11} _\Lambda {\rm B}$ and $^{12} _\Lambda {\rm C}$ taken from 
Ref.~\protect\cite{Parreno:2003ny,Parreno:2003mf}.
The theoretical numbers correspond to including WOPE, WOKE along with the
leading-order parity-violating and parity-conserving $\Lambda NNN$ interactions
with the short-distance parameters globally fit to the data.
}
\label{table:hyper}
\newcommand{\m}{\hphantom{$-$}}
\newcommand{\cc}[1]{\multicolumn{1}{c}{#1}}
\renewcommand{\tabcolsep}{1pc} 
\renewcommand{\arraystretch}{1.2} 
\begin{tabular}{@{}c|c|c}
\hline
Observable & Theory (best global fit)~\protect\cite{Parreno:2003ny,Parreno:2003mf} & Expt \\
\hline
$\Gamma ( ^5 _\Lambda {\rm He})$ & 0.44 & $0.41\pm 0.14$ , $0.50\pm 0.07$ \\
$n/p\ ( ^5 _\Lambda {\rm He})$ & 0.55 & $0.93\pm 0.55$ , $0.50\pm 0.10$ \\
${\cal A}( ^5 _\Lambda {\rm He})$ & 0.24 & $0.24\pm 0.22$ \\
\hline
$\Gamma (^{11} _\Lambda {\rm B})$ & 0.88 & $0.95\pm 0.14$\\
$n/p\ ( ^{11} _\Lambda {\rm B})$ & 0.92 & $1.04^{+0.59}_{-0.48}$ \\
${\cal A}( ^{11} _\Lambda {\rm B})$ & 0.09 & $-0.20\pm 0.10$ \\
\hline
$\Gamma (^{12} _\Lambda {\rm C})$ & 0.93 & $1.14\pm 0.2$ , $0.89\pm 0.15$ ,  $0.83\pm 0.11$\\
$n/p\ (^{12} _\Lambda {\rm C})$ & 0.77 & $0.87\pm 0.23$ \\
${\cal A}(^{12} _\Lambda {\rm C})$ & 0.03 & $-0.01\pm 0.10$ \\
\hline
\end{tabular}
\end{table}

It is found that the overall goodness of the fit encapsulated in the
$\chi^2$ is independent of the range of the short-distance Gaussian
regulator (over the range explored) and also independent of the
strong-interaction potential used to compute the final-state
interactions.  This is very encouraging.  The numerical values of the
bare coefficients in the weak potential depend upon the range of the
Gaussian regulator, as one would expect, but the dependence of the
observables is minimal.  Therefore, it appears that a systematic EFT
description of the weak decays should be possible in the near future,
and it may be as simple as Weinberg power-counting in many of the decay
channels. Clearly, this exciting possibility requires further exploration.

It goes without saying that it would be very nice to see a
calculation of the weak $\Lambda N\rightarrow NN$ amplitude directly
from lattice QCD (a 5-point function!).  While one knows how to
extract the weak decay amplitudes for $K\rightarrow \pi\pi$ from the
lattice, part of which is tuning the lattice volume so that one of the
$\pi\pi$ energy states on the lattice is degenerate with a kaon at rest,
no such formulation currently exists for extracting weak amplitudes
for $2\rightarrow 2$ processes.  In principle we do not see any formal
road blocks to constructing a framework analogous to that of
L{\"u}scher and Lellouch, however, given that the one-body decays are
not yet understood we do not see any urgency in developing such a
framework.

\section{Conclusions}

\noindent We have explored some aspects of strong-interaction physics
that impact our understanding of the structure and decays of
hypernuclei and that could be addressed by lattice QCD simulations in
the near future. In the few-nucleon sector, lattice QCD will provide
an understanding of the strong dynamics, such as the fine-tuning that
exists in the two-nucleon system, but it will not improve the
precision of input into nuclear calculations of many-body observables.
However, in the hypernuclear sector where experimental measurements
are scarce and quite uncertain, lattice QCD may be in a position to
compete with and ultimately surpass experiment and thus greatly aid
progress in the field.  We have discussed quantities in both the
strong and weak sectors whose calculation in lattice QCD would have
great significance.

To summarize, the lattice QCD calculations of strong-interaction processes that would be of fundamental importance
for hypernuclear physics are:
\begin{enumerate}
\item[$\bullet$]
An extraction of the scattering length and effective range for 
$\Lambda N$ scattering from finite-volume simulations;
\item[$\bullet$]
A determination of matrix elements of the axial current for 
$\Sigma\rightarrow\Sigma$, $g_{\Sigma\Sigma}$;
\item[$\bullet$]
A determination of the $\eta NN$ and $\eta\Lambda\Lambda$ hairpin couplings in PQQCD;
\end{enumerate}
and of weak-interaction processes are:
\begin{enumerate}
\item[$\bullet$]
A measurement of the coefficients of the weak nonleptonic two-point functions, $h_\Lambda$ and $h_\Sigma$;
\item[$\bullet$]
A measurement of both the $S-$ and $P-$wave amplitudes for the weak nonleptonic decays,
e.g. $\Lambda\rightarrow p\pi$;
\item[$\bullet$]
A measurement of the amplitude for the weak scattering $\Lambda N\rightarrow NN$.
\end{enumerate}

A lattice QCD determination of any one of these quantities would significantly improve our
understanding of hyperons, and hypernuclei in general.

\bigskip\bigskip


\acknowledgments

\noindent We would like to thank Will Detmold, Elizabeth Jenkins,
David Lin and Aneesh Manohar for helpful discussions and SRB, AP and
MJS would like to thank Paulo Bedaque for organizing the {\it
Effective Summer in Berkeley} workshop at LBL this past summer where
this work was initiated. We also thank Rob Timmermans for a
very-detailed analysis and discussion of the scattering parameters for
hyperon-nucleon scattering.  The work of SRB was partly supported by
DOE contract DE-AC05-84ER40150, under which the Southeastern
Universities Research Association (SURA) operates the Thomas Jefferson
National Accelerator Facility. PFB was supported by the Director,
Office of Energy Research, Office of High Energy and Nuclear Physics,
and by the Office of Basic Energy Sciences, Division of Nuclear
Sciences, of the U.S.~Department of Energy under Contract
No.~DE-AC03-76SF00098.  MJS is supported in part by the U.S.~Dept. of
Energy under Grant No.~DE-FG03-97ER4014.  AP is supported by the MCyT
under Grant No.~DGICYT BFM2002--01868 and by the Generalitat de
Catalunya under Grant No.~SGR2001--64.

\section{Appendix I : L{\"u}scher's Relations from $\nopi$}

\noindent From the discussions in Refs.~\cite{Kaplan:1998tg,Kaplan:1998we} it is clear
that in the pionless theory describing non-relativistic baryons, each of mass $M$,
the exact two-body elastic scattering amplitude in the continuum is
\begin{eqnarray}
{\cal A} & = & { \sum C_{2n}p^{2n}  \over
1 - I_0 \sum C_{2n}p^{2n}}
\ \ ,\ \ 
I_0 \ = \ \left({\mu\over 2}\right)^{4-D} \int {d^{D-1}{\bf q}\over
  (2\pi)^{D-1}}
{1\over E-{|{\bf q}|^2\over M} + i \epsilon}
\ \ \ ,
\end{eqnarray}
where the $C_{2n}$ are the coefficients of operators with $2 n$ derivatives
acting on the nucleon fields (or equivalently with $n$ time derivatives).
Applying the PDS scheme to $I_0$ gives
\begin{eqnarray}
I_0^{(PDS)} & = & - {M\over 4\pi}\ \left( \mu + i p \right)\ +\ {\cal O}(D-4)
\ \ \ ,
\end{eqnarray}
where $p=\sqrt{ M E}$ and hence 
\begin{eqnarray}
{\cal A} & = & {4\pi\over M}\ {1\over p\cot\delta - i p }
\ \ \ ,
\end{eqnarray}
where the subtraction-scale dependence of the one-loop diagrams is exactly
compensated by the corresponding dependence of the coefficients $C_{2n} (\mu)$.
$\delta$ is the energy-dependent $S-$wave phase shift (we will only consider
$S-$wave scattering but this construction generalizes to all partial waves).

We are interested in the energy-eigenvalues of this system placed in a box with
sides of length $L$ with periodic boundary conditions.
We can find the energy-eigenvalues by requiring that the real part of the inverse
scattering amplitude computed in the box vanishes,
\begin{eqnarray}
{1\over \sum C_{2n}(\mu) \ p^{2n}}\ -\ {\rm Re}( I_{0}^{(PDS)}(L) ) & = & 0
\ \ \ ,
\label{eq:reamp}
\end{eqnarray}
where 
\begin{eqnarray}
I_{0}(L)& = & 
{1\over L^3} \sum_{{\bf k}} {1\over E-{|{\bf k}|^2\over M} }
\ \ \ ,
\end{eqnarray}
and the sum is over momenta, ${\bf k}$, allowed on the lattice.

The PDS value of this linearly-divergent integral is found by adding
and subtracting the continuum limit of the integral evaluated at
$E=0$.  One of the continuum integrals is evaluated with a momentum
cut-off that is equal to the mode cut-off of the discrete summation,
while the other is evaluated with dimensional-regularization and PDS.
The subtraction-scheme dependence vanishes as expected, and we find
that eq.~(\ref{eq:reamp}) becomes
\begin{eqnarray}
&& p\cot\delta(p) 
\ -\ 
{1\over \pi L} \sum_{{\bf j}}^{\Lambda_j} 
{1\over |{\bf j}|^2-({Lp\over 2\pi})^2}
\ +\ 
{4 \Lambda_j\over L}
\ = \ 0
\ \ \ ,
\label{eq:energies}
\end{eqnarray}
where the limit $\Lambda_j\rightarrow\infty$ is understood, and where
$\Lambda_j$ is the magnitude of the integer cut-off, related to the
momentum cut-off in the box defined via $|{\bf k}_{\rm max.}|= 2\pi
\Lambda_j/L$.  Solving for the values of $p$ for which
eq.~(\ref{eq:energies}) is satisfied recovers L{\"u}schers
result(s)~\cite{Luscher:1986pf,Luscher:1990ux}.  Given that
$\Lambda_j$ is set by the edge of the first Brillouin zone, clearly
one must not be too cavalier about taking the
$\Lambda_j\rightarrow\infty$ limit in present day
calculations~\footnote{This issue is discussed in Ref.~\cite{DavidL}. We thank William Detmold and David Lin for
discussions regarding this point.}, however, in this work we will take
the limit $\Lambda_j\rightarrow\infty$.  This derivation of
L{\"u}scher's formula~\cite{Luscher:1986pf,Luscher:1990ux} makes
explicit the analytic continuation of the zeta-functions that appear
in his expressions.

Using the effective-range expansion of $p\cot\delta$
\begin{eqnarray}
p\cot\delta & = & 
-{1\over a}\  +\  {1\over 2} r p^2 \ +\  ...
\ \ \ ,
\label{eq:erexp}
\end{eqnarray}
and assuming that $L\gg a, r$~\footnote{It is interesting to note that
  one can also use this expression to find the energy-eigenvalues in
  the limit where $r\ll L\ll a$, as may be appropriate for two- or
  few-body nuclear physics observables.  We will explore this limit
  in forthcoming work.}  
one finds that the lowest-energy eigenvalue is
\begin{eqnarray}
E_0 & = & + {4\pi a\over M L^3}\left[\ 1\ -\ c_1 {a\over L}\ 
+\ c_2 \left({a\over L}\right)^2\ +\ ...\right]
\ +\ {\cal O}(L^{-6})
\ \ \ ,
\label{eq:e0}
\end{eqnarray}
where the coefficients $c_{1,2}$ are 
\begin{eqnarray}
c_1 & = & 
{1\over \pi} \left(\ \sum_{{\bf j}\ne 0}^{\Lambda_j} {1\over |{\bf
      j}|^2} - 4\pi\Lambda_j\ \right)
\ =\ 
-2.837297
\ =\ {1\over \pi} Z_{00}(1,0) \ ;
\nonumber\\
c_2 & = & c_1^2 - {1\over\pi^2}\sum_{{\bf j}\ne {\bf 0}} {1\over |{\bf j}|^4}
\ =\ 
6.375183
\ = \ 
{1\over\pi^2}\left(\ 
(Z_{00}(1,0))^2 -  Z_{00}(2,0)\ \right)
\ \ \ ,
\end{eqnarray}
where the $Z_{00}(n,k)$ are defined in
Refs.~\cite{Luscher:1986pf,Luscher:1990ux},
\begin{eqnarray}
Z_{00}(n,k) & = & \lim_{\Lambda_j\rightarrow\infty}
\left(\ 
\sum_{{\bf j} ,  |{\bf j}|\ne k}^{\Lambda_j}
{1\over ( |{\bf j}|^2 - k^2)^n}
\ -\ 
\delta^{n1}\ 4\pi\Lambda_j
\ \right)
\ \ \ ,
\end{eqnarray}
for integers $n\geq 1$ and $k$.  It is important to note that there are sign
differences between our result in eq.~(\ref{eq:e0}) and the analogous
expression in Ref.~\cite{Luscher:1986pf,Luscher:1990ux}.  This is due
to the opposite sign convention for the scattering length $a$, for
which we use that common to nuclear physicists.

We also need the energy of the next state in order to extract both the
scattering length and the effective range.  The construction is
somewhat more complicated than that for the lowest state.  One looks
at the energy of the next-most energetic level that is invariant under
the action of the complete cubic group; the space of such states is
the $A_1$ sector~\cite{Luscher:1986pf,Luscher:1990ux}.  It is easy to
show from the decomposition of the full rotation group down to the
cubic group~\cite{Mandula:ut} that the energy shift of such states due
to the two-body interaction receives contributions from $l=0, 4, 6,
..$ interactions, where $l$ is the angular momentum of the
interaction.  Therefore, to the order we are working in the momentum
expansion, the shift in the energy-levels is sensitive to $S-$wave
interactions only.  It is straightforward to show that the energy of
the next-highest state in the $A_1$ representation of the cubic group
is~\cite{Luscher:1986pf,Luscher:1990ux}
\begin{eqnarray}
E_1 & = & {4\pi^2\over M L^2} - 
{12\tan\delta_0\over M L^2}\left(\ 
1 + c_1^\prime\tan\delta_0 + c_2^\prime \tan^2\delta_0\ +\ ...\ \right)
\ +\ {\cal O}(L^{-6}) \ ;
\nonumber\\ 
c_1^\prime & = & 
{1\over 2\pi^2} Z_{00}(1,1) \ ;
\nonumber\\ 
& = & {1\over 2\pi^2}\left[\ 
\pi c_1 + \pi^2 c_1^2 -\pi^2 c_2 -13 + \sum_{{\bf j} , |{\bf j}|\ne 0,1}^{\Lambda_j} {1\over
|{\bf j}|^4 ( |{\bf j}|^2-1 )}
\right]
\ =\ -0.061367 \ ;
\nonumber\\ 
c_2^\prime & = &  
{1\over 4\pi^4}\left(\ (Z_{00}(1,1))^2 - 6 Z_{00}(2,1)\
\right) \ ;
\nonumber\\ 
& = &
c_1^{\prime 2}
 + {3\over 2\pi^4}\left(\ 
\pi^2 c_2 - \pi^2 c_1^2 + 5 - \sum_{ {\bf j} ,|{\bf j}|\ne 0,1}^{\Lambda_j} 
{2 |{\bf j}|^2 -1\over |{\bf j}|^4 ( |{\bf j}|^2-1 )^2}
\ \right)
\ =\ -0.354156
\ \ \ ,
\label{eq:e1}
\end{eqnarray}
where $\delta_0$ is the $S-$wave phase shift evaluated at the unperturbed
lattice momentum, $|{\bf p}| = 2\pi/L$, and coefficients $c_{1,2}^\prime$ are
those found by L{\"u}scher~\cite{Luscher:1986pf,Luscher:1990ux}.
We have given expressions for $c_{1,2}^\prime$ in terms of $c_{1,2}$ that
can be straightforwardly evaluated and converge rapidly.

To generalize the expressions in eq.~(\ref{eq:e0}) and
eq.~(\ref{eq:e1}) to $\Lambda N$ scattering we make the replacement
$M\rightarrow 2 \mu_{\Lambda N}$ and understand that $\delta_0$ is the
phase shift for $\Lambda N$ scattering with the scattering lengths and
effective ranges discussed in the text evaluated at the unperturbed
momentum $|{\bf p}| = 2\pi/L$.



\begin{thebibliography}{100}

\bibitem{exp} For a review and comparative study of the different reaction 
mechanisms, see W.~Alberico and G.~Garbarino, 
{\it Phys. Rept.} {\bf 369}, 1 (2002),
{\tt nucl-th/0112036}.

\bibitem{heliexp}
J.J.~Szymanski {\it et al.}, 
{\it Phys. Rev.} {\bf C43}, 849 (1991);
H.~Noumi {\it et al.}, {\it Phys. Rev.} {\bf C52}, 2936 (1995);
H.~Outa {\it et al.}, talk presented at the VIIIth Intl.~Conf.~on Hypernuclear
and Strange Particle Physics (HYP03), Jefferson Lab, Newport News, Virginia, 
Oct. 14-18, 2003, To be published in {\it Nucl. Phys.} {\bf A}.

\bibitem{hypertriton}
J.~Golak, K.~Miyagawa, H.~Kamada, H.~Witala, W.~Gl\"ockle, A.~Parre\~no, 
A.~Ramos, and C.~Bennhold, 
{\it Phys. Rev.} {\bf C55}, 2196 (1997); {\it ibid.} 
{\it Phys. Rev.} {\bf C56}, 2892 (1997).

\bibitem{nogga}
A.~Nogga, H.~Kamada, and W.~Gl\"ockle, 
{\it Phys. Rev. Lett.} {\bf 88}, 172501 (2002),
{\tt nucl-th/0112060}.

\bibitem{hiyama}
E.~Hiyama, Y.~Kino, M.~Kamimura, 
{\it Prog. Part. Nucl. Phys.} {\bf 51}, 223 (2003).

\bibitem{rob}
C.L.~Korpa, A.E.L.~Dieperink and R.G.E.~Timmermans, 
{\it Phys. Rev.}  {\bf C65}, 015208 (2001),
{\tt nucl-th/0109072}.

\bibitem{DFHT96}
J.F.~Dubach, G.B.~Feldman and B.R.~Holstein,
{\it Annals Phys.} {\bf 249}, 146 (1996),
{\tt nucl-th/9606003}.

\bibitem{OMEW}
A.~Parre\~no, A.~Ramos, and C.~Bennhold, 
{\it Phys. Rev.} {\bf C56}, 339 (1997),{\tt nucl-th/9611030};
A.~Parre\~no and A.~Ramos, 
{\it Phys. Rev.} {\bf C65} (2002) 015204,{\tt nucl-th/0104080}.

\bibitem{ISO00}
T.~Inoue, K.~Sasaki, M.~Oka, 
{\it Nucl. Phys.} {\bf A670}, 301 (2000).

\bibitem{Parreno:2003ny}
A.~Parre\~no, C.~Bennhold and B.R.~Holstein,
{\tt nucl-th/0308074}.

\bibitem{Parreno:2003mf}
A.~Parre\~no, C.~Bennhold and B.R.~Holstein,
{\tt nucl-th/0308056}.


\bibitem{Fukugita}
M.~Fukugita, Y.~Kuramashi, M.~Okawa, H.~Mino and A.~Ukawa,
{\it Phys. Rev. Lett.}  {\bf 73}, 2176 (1994),
{\tt hep-lat/9407012};
{\it Phys. Rev.} {\bf D52}, 3003 (1995),
{\tt hep-lat/9501024}.

\bibitem{NNonline}
{\tt http://nn-online.sci.kun.nl/index.html}.

\bibitem{Sechi-Zorn:hk}
B.~Sechi-Zorn, B.~Kehoe, J.~Twitty and R.A.~Burnstein,
{\it Phys. Rev.} {\bf 175}, 1735 (1968).

\bibitem{Alexander:cx}
G.~Alexander, U.~Karshon, A.~Shapira, G.~Yekutieli, R.~Engelmann, H.~Filthuth and W.~Lughofer,
{\it Phys. Rev.}  {\bf 173}, 1452 (1968).

\bibitem{hyperexamples}
See, for instance, 
Y.C.~Tang, in {\it Proc. of the Int. Conf. on Hypernuclear Physics},
Argonne Nat. Lab. (Ed. A.R.~Bodmer and L.G.~Hyman), p. 276;
R.C.~Herndon and Y.C.~Tang, {\it Phys. Rev.} {\bf 159}, 835 (1967);
R.H.~Dalitz, R.C.~Herndon and Y.C.~Tang, {\it Nucl. Phys.} {\bf B47}, 109 (1972);
T.H.~Tan, {\it Phys. Rev. Lett.} {\bf 23}, 395 (1969);
J.J.~de Swart {\it et al.}, {\it Springer Tracts in Modern Physics} {\bf 60}, 138 (1971).

\bibitem{Weinberg:rz}
S.~Weinberg,
{\it Phys. Lett.} {\bf B251}, 288 (1990).

\bibitem{Weinberg:um}
S.~Weinberg,
{\it Nucl. Phys.} {\bf B363}, 3 (1991).

\bibitem{Ordonez:1992xp}
C.~Ordonez and U.~van Kolck,
{\it Phys. Lett.} {\bf B291}, 459 (1992).

\bibitem{Ordonez:tn}
C.~Ordonez, L.~Ray and U.~van Kolck,
{\it Phys. Rev. Lett.}  {\bf 72}, 1982 (1994).

\bibitem{Kaplan:1996xu}
D.B.~Kaplan, M.J.~Savage and M.B.~Wise,
{\it Nucl. Phys.} {\bf B478}, 629 (1996),
{\tt nucl-th/9605002}.

\bibitem{Kaplan:1998tg}
D.B.~Kaplan, M.J.~Savage and M.B.~Wise,
{\it Phys. Lett.} {\bf B424}, 390 (1998),
{\tt nucl-th/9801034}.

\bibitem{Kaplan:1998we}
D.B.~Kaplan, M.J.~Savage and M.B.~Wise,
{\it Nucl. Phys.}  {\bf B534}, 329 (1998),
{\tt nucl-th/9802075}.

\bibitem{Kaplan:1998sz}
D.B.~Kaplan, M.J.~Savage and M.B.~Wise,
{\it Phys. Rev.} {\bf C59}, 617 (1999),
{\tt nucl-th/9804032}.

\bibitem{Beane:2001bc}
S.R.~Beane, P.F.~Bedaque, M.J.~Savage and U.~van Kolck,
{\it Nucl. Phys.} {\bf A700}, 377 (2002),
{\tt nucl-th/0104030}.

\bibitem{Birse}
M.C.~Birse, J.A.~McGovern and K.G.~Richardson,
{\it Phys. Lett.} {\bf B464}, 169 (1999),
{\tt hep-ph/9807302}.

\bibitem{Luscher:1986pf}
M.~L{\"u}scher,
{\it Commun. Math. Phys.}  {\bf 105} 153 (1986).

\bibitem{Luscher:1990ux}
M.~L{\"u}scher,
{\it Nucl. Phys.} {\bf B354}, 531 (1991).

\bibitem{Hammer:2001ng}
H.W.~Hammer,
{\it Nucl. Phys.} {\bf A705}, 173 (2002),
{\tt nucl-th/0110031}.


\bibitem{Chen:2001yi}
J.W.~Chen and M.J.~Savage,
{\it Phys. Rev.} {\bf D65}, 094001 (2002),
{\tt hep-lat/0111050}.

\bibitem{Beane:2002vq}
S.R.~Beane and M.J.~Savage,
{\it Nucl. Phys.}  {\bf A709}, 319 (2002),
{\tt hep-lat/0203003}.

\bibitem{Beane:2002ca}
S.R.~Beane and M.J.~Savage,
{\it Nucl. Phys.} {\bf B636}, 291 (2002),
{\tt hep-lat/0203028}.

\bibitem{Beane:2002nu}
S.R.~Beane and M.J.~Savage,
{\it Phys. Lett.} {\bf B535}, 177 (2002),
{\tt hep-lat/0202013}.

\bibitem{Beane:2002np}
S.R.~Beane and M.J.~Savage,
{\it Phys. Rev.}  {\bf D67}, 054502 (2003),
{\tt hep-lat/0210046}.

\bibitem{Arndt:2003vx}
D.~Arndt, S.R.~Beane and M.J.~Savage,
{\it Nucl. Phys.} {\bf A726}, 339 (2003),
{\tt nucl-th/0304004}.

\bibitem{Cho:1992cf}
P.L.~Cho,
{\it Nucl. Phys.} {\bf B396}, 183 (1993);
[Erratum-ibid.\ {\bf B421}, 683 (1994)],
{\tt hep-ph/9208244}.

\bibitem{Aneesh}
R.F.~Dashen, E.~Jenkins and A.V.~Manohar,
{\it Phys. Rev.} {\bf D49}, 4713 (1994),
[Erratum-ibid.\ {\bf D51}, 2489 (1995)],
{\tt hep-ph/9310379}.

\bibitem{Kaplan:1995yg}
D.B.~Kaplan and M.J.~Savage,
{\it Phys. Lett.} {\bf B365}, 244 (1996),
{\tt hep-ph/9509371}.

\bibitem{Sharpe:2001fh}
S.R.~Sharpe and N.~Shoresh,
{\it Phys. Rev.}  {\bf D64}, 114510 (2001),
{\tt hep-lat/0108003}.


\bibitem{Maiani:ca}
L.~Maiani and M.~Testa,
{\it Phys. Lett.} {\bf B245}, 585 (1990).

\bibitem{yang}
K.~Huang and C.N.~Yang,
{\it Phys. Rev.} {\bf 105}, 767 (1957).

\bibitem{Chen:1999tn}
J.W.~Chen, G.~Rupak and M.J.~Savage,
{\it Nucl. Phys.}  {\bf A653}, 386 (1999),
{\tt nucl-th/9902056}.

\bibitem{vanKolck:1998bw}
U.~van Kolck,
{\it Nucl. Phys.}  {\bf A645}, 273 (1999),
{\tt nucl-th/9808007}.

\bibitem{vanBaal:2000zc}
P.~van Baal,
In the Boris Ioffe Festschrift, 
{\it At the frontier of particle physics}, vol. 2, 683-760, 
ed. by M. Shifman, World Scientific,
{\tt hep-ph/0008206}.


\bibitem{Donoghue:dd}
J.F.~Donoghue, E.~Golowich and B.R.~Holstein,
{\it Cambridge Monogr. Part. Phys. Nucl. Phys. Cosmol.}  {\bf 2}, 1 (1992).

\bibitem{Jenkins:1991bt}
E.~Jenkins,
{\it Nucl. Phys.} {\bf B375}, 561 (1992).

\bibitem{Jenkins:1991ne}
E.~Jenkins and A.V.~Manohar,
UCSD-PTH-91-30,
{\it Talk presented at the Workshop on Effective Field Theories of the Standard Model, Dobogoko, Hungary, Aug 1991}.

\bibitem{Savage:1996zd}
M.J.~Savage and J.~Walden,
{\it Phys. Rev.}  {\bf D55}, 5376 (1997),
{\tt hep-ph/9611210}.

\bibitem{Springer:1995nz}
R.P.~Springer,
{\tt hep-ph/9508324}.

\bibitem{Springer:sv}
R.P.~Springer,
{\it Phys. Lett.} {\bf B461} (1999) 167.

\bibitem{AbdEl-Hady:1999mj}
A.~Abd El-Hady and J.~Tandean,
{\it Phys. Rev.} {\bf D61}, 114014 (2000),
{\tt hep-ph/9908498}.

\bibitem{Dai:1995zg}
J.~Dai, R.F.~Dashen, E.~Jenkins and A.V.~Manohar,
{\it Phys. Rev.} {\bf D53}, 273 (1996),
{\tt hep-ph/9506273}.

\bibitem{Flores-Mendieta:1998ii}
R.~Flores-Mendieta, E.~Jenkins and A.V.~Manohar,
{\it Phys. Rev.} {\bf D58}, 094028 (1998),
{\tt hep-ph/9805416}.


\bibitem{Lellouch:2000pv}
L.~Lellouch and M.~L{\"u}scher,
{\it Commun. Math. Phys.}  {\bf 219}, 31 (2001),
{\tt hep-lat/0003023}.


\bibitem{DavidL}
C.J.~Lin, G.~Martinelli, C.T.~Sachrajda and M.~Testa, 
{\it Nucl. Phys.} {\bf B619}, 467 (2001),
{\tt hep-lat/0104006}. 

\bibitem{Mandula:ut}
J.E.~Mandula, G.~Zweig and J.~Govaerts,
{\it Nucl. Phys.} {\bf B228}, 91 (1983).

\end{thebibliography}
\end{document}